\documentclass[prb,twocolumn,showpacs,amsmath,amssymb,superscriptaddress,floatfix,reprint,nofootinbib]{revtex4-2}
\usepackage{amssymb}
\usepackage{amsbsy}
\usepackage{amsmath}
\usepackage{mathtools}
\usepackage{mathrsfs}
\usepackage{epsfig}
\usepackage{graphicx}
\usepackage{array}
\usepackage{textcomp}
\usepackage{xcolor}
\usepackage{braket}
\usepackage{hhline}
\usepackage{dcolumn}
\usepackage{dsfont}
\usepackage{comment}
\usepackage{bm,hyperref}
\usepackage{algpseudocode}
\usepackage{algorithm}
\usepackage[justification=raggedright,font=small]{caption}
\usepackage[titletoc,toc,title]{appendix}
\usepackage[normalem]{ulem}
\usepackage[caption=false]{subfig}

\hypersetup{
	colorlinks=true,
	linkcolor=blue,      
    citecolor=blue,
}

\newcommand{\tr}{\text{Tr}}
\newcommand{\eq}[2]{\begin{equation} \label{#1} #2 \end{equation}}
\newcommand{\eqsp}[1]{\begin{split} #1 \end{split}}

\newcommand{\be}{\begin{equation}}
\newcommand{\ee}{\end{equation}}
\newcommand{\iea}{\begin{equation}\begin{aligned}}
\newcommand{\fea}{\end{aligned}\end{equation}}

\newcommand{\mbZ}{\mathbb{Z}}

\newcommand{\mE}{\mathcal{E}}
\newcommand{\eval}[1]{\langle #1 \rangle}

\begin{document}

\author{Sanket Chirame}
\affiliation{School of Physics and Astronomy, University of Minnesota, Minneapolis, Minnesota 55455, USA}
\author{Fiona J. Burnell}
\affiliation{School of Physics and Astronomy, University of Minnesota, Minneapolis, Minnesota 55455, USA}
\author{Sarang Gopalakrishnan}
\affiliation{Department of Electrical and Computer Engineering, Princeton University, Princeton, NJ 08544}
\author{Abhinav Prem}
\affiliation{School of Natural Sciences, Institute for Advanced Study, Princeton, New Jersey 08540, USA}

\title{Stable Symmetry-Protected Topological Phases in Systems with Heralded Noise}

\date{\today}

\begin{abstract}
    We present a family of local quantum channels whose steady-states exhibit stable mixed-state symmetry-protected topological (SPT) order. Motivated by recent experimental progress on ``erasure conversion'' techniques that allow one to identify (\emph{herald}) decoherence processes, we consider open systems with biased erasure noise, which leads to strongly symmetric heralded errors. We utilize this heralding to construct a local correction protocol that effectively confines errors into short-ranged pairs in the steady-state. Using a combination of numerical simulations and mean-field analysis, we show that our protocol stabilizes SPT order against a sufficiently low rate of decoherence. As the rate of heralded noise increases, SPT order is eventually lost through a directed percolation transition. We further find that while introducing unheralded errors destroys SPT order in the limit of long length- and time-scales, the correction protocol is sufficient for ensuring that local SPT order persists, with a correlation length that diverges as $\xi \sim (1-f_e)^{-1/2}$, where $f_e$ is the fraction of errors that are heralded.
\end{abstract}

\maketitle


\begin{figure}
\subfloat{%
  \includegraphics[width=.95\columnwidth]{{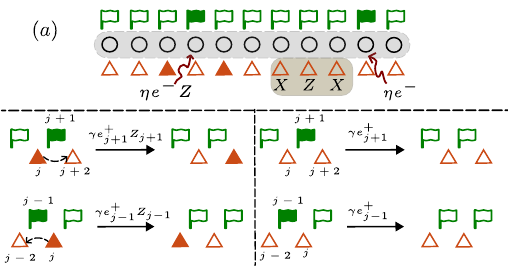}}%
}\hfill 
\\ 
\subfloat{%
  \includegraphics[width=.95\columnwidth]{{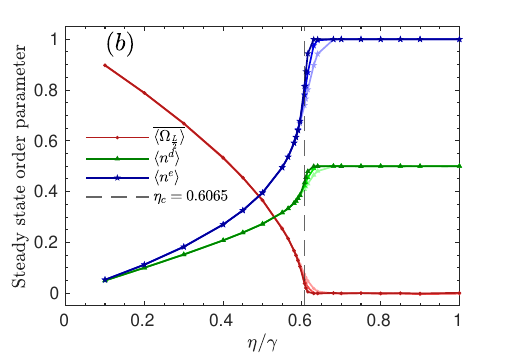}}%
}\hfill
\caption{\label{mainfig-1}(a) Top: schematic showing stabilizers $X_{j-1} Z_j X_{j+1}$ (triangles) and locations of erasure errors (flags) relative to the qubits (circles). Bottom: the correction protocol for different configurations (the remainder of the system is unchanged). The left displays processes that move a stabilizer defect (filled triangle) past an adjacent erasure (filled flag); the right shows processes that shorten erasure strings which do not end on a stabilizer defect. (b) The steady-state expectation values of the string-order ($\Omega$), density of stabilizer defects ($n^d$), and the density of erasures ($n^e$) are shown as a function of the erasure noise rate $\eta/\gamma$ and clearly display a sharp transition. The data is shown for  $L=128,256,512$ sites per sublattice (increasing system sizes from lighter to darker shades). The steady-state values are obtained by averaging over $5-10\times 10^3$ independent Monte Carlo trajectories initialized in the cluster state and simulated upto $2\times L^{1.5}$ MC sweeps.}
\end{figure}


\emph{Introduction}.--- 
Quantum states with structured entanglement can serve as resources for quantum information processing tasks, such as quantum teleportation or quantum computation. In the case of ground states of local Hamiltonians, it is understood that this utility is a property of \emph{phases}: perturbing a resource state's parent Hamiltonian or applying a finite-time unitary evolution to it simply generates a different resource state, provided the perturbations respect the relevant symmetries~\cite{bravyi2010,bravyi2011short,michalakis2013}. Whether resource states can be stabilized against noise, decoherence, or continuous projective measurements is much less clear~\cite{hseih2023mixed,khemani2024absorbing,morales2023steering,chen2023steering}. In sufficiently high dimensions, phenomena such as topological order persist at nonzero temperature, as properties of the \emph{mixed} thermal density matrix~\cite{hastings2011finiteT,roberts2017finiteT,lu2020negativity,stahl2021finiteT,lu2023surface}. Even in the absence of thermal stability, ordered states can be stabilized by active quantum error correction. However, active error correction requires inherently nonlocal classical processing---identifying the recovery operation requires global knowledge of the measurement outcomes on error syndromes~\cite{terhalreview}. Requiring such nonlocal operations limits the scalability of the process \emph{in principle}; moreover, breaking spatial locality (even classically) leads to violations of the Lieb-Robinson bound and leads to effects like phase transitions in finite-depth circuits, which are absent in fully local models. In light of these considerations, it is desirable to explore how far error correction can be done \emph{locally}. 

A particularly important goal is obtaining a \textit{steady-state phase} within which each mixed state can be leveraged for quantum information tasks (see Refs.~\cite{sang2023mixed,ma2023avg,rakovszky2023stable,sang2024} for notions of phases in open quantum systems) or from which the original pure state can be recovered via a quasi-local finite depth quantum channel~\cite{sang2023mixed}. 
This task has received considerable attention in the search for self-correcting quantum memories~\cite{KitaevToric,bombin2013scm,Pastawski2014} in the context of low-dimensional topological order, where steady-state phases can be stabilized by effectively engineering long-ranged interactions between defect pairs created by noise ~\cite{harringtonPhD,Hamma2009,Chesi2010,DauphinaisPoulin,Fuji2014,herold2015,herold2017,vasmer2020}; however, while more local than conventional error correction, these models are neither strictly local nor stable to generic perturbations~\cite{Landon2015}.
Requiring stability also rules out strategies that target desired pure states as dark states of local Lindblad evolution, which are generically unstable (unless the targeted state is stable to finite temperature e.g., the 4D toric code)~\cite{kraus2008preparation,verstraete2009,lang2015,lieu2020,jamadagni2022,liu2024,wang2023}.

In this work, we show that for a particular class of experimentally relevant error models, {\it strictly local} error correction can stabilize one-dimensional symmetry protected topological (SPT) order in steady-state phases. These steady-state phases have a non-vanishing  string-order parameter, ensuring that the mixed state remains a resource for quantum teleportation~\cite{raussendorf2023} and measurement-based quantum computation (MBQC)~\cite{gottesmanMBQC,raussendorfMBQC}. In closed quantum systems, SPT order is stable against symmetric local perturbations, for which symmetry defects are locally created in pairs.  
In open quantum systems, only ``strong" symmetries correspond to the conservation of symmetry charges within the system~\cite{albert2014sym} and are hence required for defining mixed-state SPT order~\footnote{Mixed-state order can be defined for $G \times H$ symmetric systems, where $G$ is a ``weak" symmetry and $H$ is strong~\cite{ma2023prx,ma2023avg,zhang2022strange,lee2022aspt,zhou2023anomaly,hsin2023anomaly,kawabata2024anomaly,lessa2024anomaly,sohal2024imto}.}~\cite{coser2019class,degroot2022og,molnar2022mpo1,molnar2022mpo2,ma2023prx,ma2023avg,zhang2022strange,lee2022aspt,guo2023cluster2d}. While pure SPT states are stable to strongly symmetric finite-depth local channels~\cite{coser2019class,degroot2022og,paszko2023}, here we demonstrate that mixed SPT phases can arise as stable \textit{steady-states} under local Lindbladian evolution.

Our approach is motivated by the recent experimental technique of erasure conversion, which can be exploited to greatly increase error correction thresholds~\cite{stace2009,whiteside2014,wu2022,kang2023,gu2023erasure,kubica2023,sahay2023}. Erasure conversion involves designing qubits such that the dominant error processes are ``heralded erasures,'' which take a qubit out of the computational manifold into a specially marked state, with the location of a potential error detected experimentally before the system is re-initialized in the computational Hilbert space.
We focus specifically on biased erasure noise, which naturally preserves the strong symmetry and thus ensures that the noise pair-creates defects. The erasure positions contain local information about which defects were pair-created by error strings; we exploit this information to construct our fully local error correction procedure. Strikingly, this protocol succeeds in stabilizing steady-state string-order (i.e., it error corrects into the SPT phase) up to a finite noise threshold, beyond which erasures proliferate in an absorbing state transition to the trivial phase (where string-order vanishes). While imperfect heralding destroys steady-state order at long distances, we show that when most errors are heralded, our protocol strongly enhances finite-distance and finite-time string-order.


\emph{Background on SPTs}.---We will focus on stabilizing the 1D cluster state $\ket{\psi}_C = \prod_{j=1}^{2L} CX_{j,j+1} \ket{\uparrow}^{\otimes 2L}$ (where $CX_{i,j}=\frac{1}{2}\left(1+X_j+X_{j+1}-X_j X_{j+1}\right)$ is a two-qubit conditional unitary), which is an SPT state protected by internal $G = \mathbb{Z}_2 \times \mathbb{Z}_2$ symmetry. $\ket{\psi}_C$ is a representative stabilizer state within this SPT phase, satisfying $S_j \ket{\psi}_C = \ket{\psi}_C$ for all $j$, with stabilizers $S_j=X_{j-1}Z_{j}X_{j+1}$. While $\ket{\psi}_C$ is short-range entangled, it cannot be connected to a $G$-symmetric paramagnetic ground state (i.e. a symmetric product state) through a $G$-symmetric finite-depth local unitary~\cite{pollmann2010SPT,chen2011SPT,chen2013SPT}. The resulting SPT order is encoded in the nonlocal ``string-order parameters," $\Omega^a_{i_0, j_0} = X_{2i_0-1} \left( \prod_{i= i_0}^{j_0} Z_{2i} \right ) X_{2 j_0 +1} $, $\Omega^b_{i_0, j_0} = X_{2i_0}\left( \prod_{i= i_0}^{j_0} Z_{2i+1} \right )  X_{2 j_0 +2}  $, with global symmetry generators $\mathbb{Z}_2^{(a)} = \prod_{i=1}^{L} Z_{2i}$,  $\mathbb{Z}_2^{(b)} = \prod_{i=1}^{L} Z_{2i-1}$ for a system with $L$ unit cells. At generic points within the phase, when the interval $(i_0, j_0)$ spans the entire system, and thus has a definite symmetry charge, the string operators reduce to long-range two-point correlation functions of operators localized at the boundaries, resulting in edge states with fractionalized symmetry (see Supplemental Material (SM)~\cite{supmat} for details).
This leads to entanglement features that are a resource for teleportation or MBQC \textit{throughout} the SPT phase~\cite{chung2009,doherty2009,else2012spt,prakash2015spt,stephen2017spt,marvian2017spt,raussendorf2023,hong2023}.


\emph{Noise Model.---}
 Heralded noise occurs in systems where errors leave a signature that can be detected without disturbing wave-functions in the logical space. Here, we consider a particular model of heralded noise, namely biased erasure noise, which naturally exhibits strong $\mbZ_2 \times \mbZ_2$ symmetry.  This model is specifically motivated by the Rydberg atom setup considered in Ref.~\cite{wu2022,sahay2023}, where qubits are formed using metastable states of neutral $^{171}$Yb atoms. In these atomic qubits single qubit, measurement, and idling errors are extremely unlikely~\cite{wu2022,maHighfidelityGatesMidcircuit2023}; we hence neglect these for now. Instead, the dominant source of errors are two-qubit gate errors that arise during state preparation or in manipulations required to utilize the state as a resource. These errors primarily stem from a decay process that takes the system out of the computational Hilbert space and constitute experimentally detectable erasure errors. These erasures can be detected without affecting the logical state of the qubit, yielding a model with heralded noise. Since these decays
only occur from one of the two computational states, re-initializing the system in this state can produce at most $Z$-errors. 

After Pauli twirling, these biased erasure errors are described by the jump operators (see SM~\cite{supmat} for details):
\eq{}{ \label{Eq:HerErr} \eqsp{
L_{\eta, 1, j}  = \sqrt{\eta/2} \, Z_j e^-_j \ , \ \ \ & L_{\eta, 2, j} = \sqrt{\eta/2} \,  Z_j n^e_j \\ 
L_{\eta, 3, j} = \sqrt{\eta/2} \, e^-_j \ , \ \ \ & L_{\eta, 4, j} = \sqrt{\eta/2} \, n^e_j.
}}
Here $e^-=|1\rangle\langle0|$ and $e^+=|0\rangle\langle1|$ are raising and lowering operators for the number $n^e_j$ of erasures on site $j$ (which lie outside the logical space), $Z_j$ is the Pauli-$Z$ operator in the logical space, and $\eta$ parametrizes the background erasure noise rate, with $\eta d t$ the error probability in a single application of the corresponding noise channel (dropping terms of order $(\eta d t)^2$). This channel has two crucial features: first, in this idealized limit, \textit{all} errors are heralded; second, it has strong $\mathbb{Z}_2 \times \mathbb{Z}_2$ symmetry.

\emph{Local error correction.}---As noted above, strong symmetry is insufficient for preserving SPT order in the steady-state. Rather, it guarantees that stabilizer violations (or `defects') are \textit{pair-created} at end-points of contiguous error strings; subsequent errors cause them to undergo diffusive dynamics, leading to a uniform density of defecets in the steady-state. Specifically, the string-order for a subregion $R$ is $1$ ($-1)$ in any pure state where the number of defects within $R$ is even (odd). For a single application of the noise channel, the probability of generating an odd number of defects in $R$ is the probability of creating a pair of defects across the subregion's boundary, which depends on the error rate $\eta$ but \textit{not} the system size: strong-symmetry thus guarantees a finite lifetime for the string-order parameter~\cite{coser2019class}. However, in the steady-state of the noise channel even and odd numbers of defects occur with equal probability: nontrivial steady-state string-order hence requires a correction channel to bind the defect pairs that occur at end-points of error strings. In the SM~\cite{supmat}, we discuss a potential strategy that enforces biased motion of defects and leads to a robust mixed-state SPT \textit{phase} with nontrivial string-order; however, this protocol is sensitive to boundary conditions. 

Here, we instead exploit the additional information provided by heralded noise to confine defect pairs, thereby stabilizing steady-state SPT order. Consider a 1D chain with $2L$ qubits and periodic boundary conditions (PBC). Each phase flip $Z_j$ error in the cluster state $\ket{\psi}_C$ creates a pair of {\it defects} $S_{j-1} = S_{j+1} = -1$. Our correction protocol is a two-step process: we first measure a syndrome to detect a defect and subsequently apply an operator that moves the defect in accordance with the configuration of proximate erasures. The direction is chosen such that pairs of defects created by the same string of contiguous errors move towards each other and ultimately annihilate.

The rules governing this motion are illustrated in Fig.~\ref{mainfig-1}a: if the syndrome for the stabilizer $S_j$ is detected, the defect is moved towards whichever neighbouring site has an erasure; the erasure flag from that site is subsequently removed. The corresponding jump operators are:
\eq{}{\label{eq:corr1}\eqsp{L_{\gamma,1,j}&=\sqrt{\gamma/4} \ e^+_{j-1}Z_{j-1} (1 - S_j)(1-n^e_{j+1}),\\
L_{\gamma,2,j}&=\sqrt{\gamma/4}\ e^+_{j+1}Z_{j+1}(1-S_j)(1-n^e_{j-1}).}}
Also, isolated erasure flags without stabilizer syndromes next to their endpoints are removed via
\eq{}{\label{eq:corr2}\eqsp{L_{\gamma,3,j}&=\sqrt{\gamma/4}\ e^+_{j-1}(1+S_j)(1-n^e_{j+1}), \\ L_{\gamma,4,j}&=\sqrt{\gamma/4}\ e^+_{j+1}(1+S_j)(1-n^e_{j-1}).}}
Here $\gamma$ is the correction rate, with a correction probability $\gamma dt$ in a single application of the correction channel $\epsilon_C(\rho)$.
A circuit implementing these correction operators is given in \cite{supmat}.


\begin{figure}
\subfloat{%
 \includegraphics[width=0.95\columnwidth]{{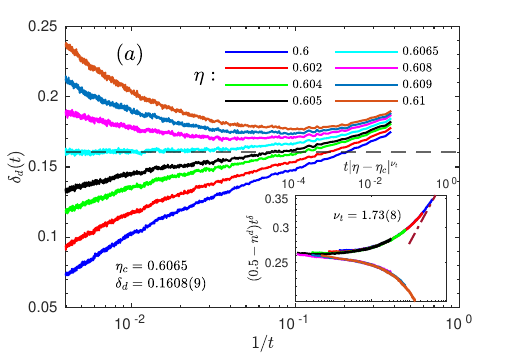}}%
}\hfill \\
\subfloat{%
  \includegraphics[width=0.95\columnwidth]{{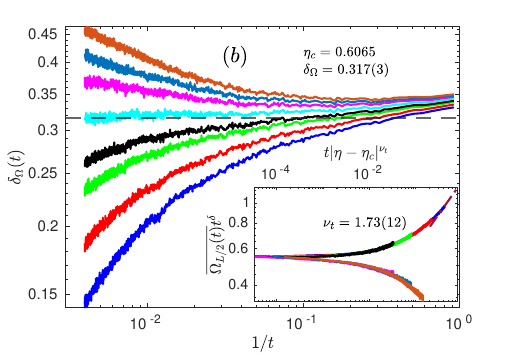}}%
}\hfill
\caption{\label{Fig:Critical} Estimation of critical exponents ($L=512$). (a) Stabilizer defects: The running estimate of the exponent $\delta(t):= \log_{10}\frac{(0.5-n^d(t))}{(0.5-n^d(10t))}$ is plotted as a function of $1/t$. The order parameter at the critical point vanishes as $t^{-\delta_d}$ so $\lim_{t\rightarrow\infty}\delta_d(t,\eta=\eta_c)=\delta_d$. In the absorbing (active) phase $\delta_d(t)\rightarrow\infty (0)$. The critical value of $\eta$ is estimated by identifying the curve that saturates to a constant value as $\frac{1}{t}\rightarrow0$ and the corresponding constant value is the exponent $\delta_d$. The inset shows the scaling collapse for $(0.5-n^d) t^\delta$ as a function of $t|\eta-\eta_c|^{\nu_t}$ after initial transient when $\nu_t$ is tuned to $1.73$. (b) The string order exponents $\delta_\Omega$ and $\nu_t$ are estimated similarly. The data is obtained by averaging over $4\times 10^4$ Monte Carlo trajectories initialized in the cluster state and simulated upto $L^{1.5}$ MC sweeps (see SM~\cite{supmat} for details).}
\end{figure}

\begin{figure}
\subfloat{%
\includegraphics[width=0.9\columnwidth]{{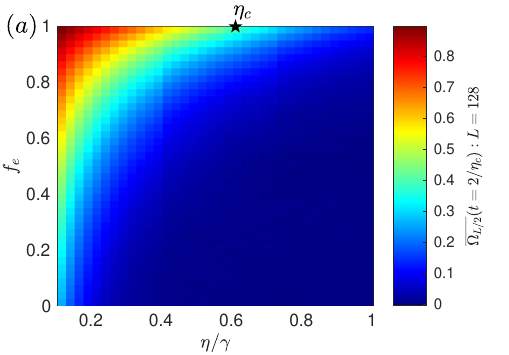}}%
}\hfill \\
\subfloat{%
  \includegraphics[width=0.5\columnwidth]{{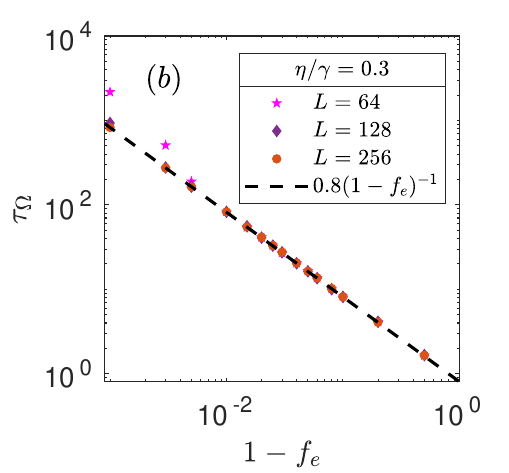}}%
}\hfill 
\subfloat{%
  \includegraphics[width=0.5\columnwidth]{{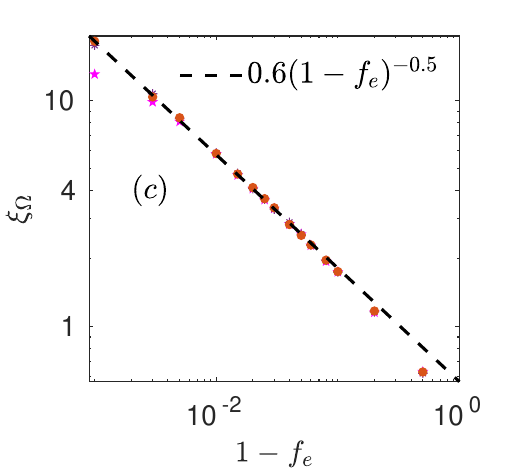}}%
}\hfill
\caption{ \label{Fig:unbiased} Finite non-erasure errors: (a) The $f_e-\eta$ phase diagram of the system. $\eta$ is the total error rate and $f_e$ is the fraction of errors that generate erasures. The color-plot shows the string-order of length $l=L/2$ evaluated at a finite time $t=2/\eta_c$, which is chosen to maximize the contrast between the active phase (where string order remains strong at this time) and the absorbing phase (where it has all but vanished at time $2/\eta_c$). This half-chain string-order decays to zero exponentially with time for any $f_e<1$ and vanishes in the steady-state. The data is obtained by averaging over $5\times 10^3$ Monte Carlo runs. (b) The corresponding time constant $\tau_\Omega$ at $\eta = 0.3< \eta_c$, obtained by a best fit to the numerical data (see \cite{supmat} for details), is plotted as a function of the non-erasure fraction $1-f_e$. The corresponding steady-state string-order is analyzed along the cut $\eta=0.3$ in (c). The characteristic length scale $\xi_\Omega$ is obtained by fitting the steady-state string-order of length $l$ to $f(l)\sim e^{-l/\xi_\Omega}$. The length scale $\xi_\Omega$ is plotted as a function of $1-f_e$ for various system sizes $L$.
}
\end{figure}


\emph{Results.---}We study the steady-state dynamics of a system subject to both biased erasure noise and local error correction. In the continuous time limit, the full time evolution of the density matrix is given by the Lindblad master equation~\cite{lindblad1976generators,gorini}:
\eq{}{ \label{Eq:Lindblad}
d\rho/dt = \sum_{\alpha = \gamma, \eta}\sum_{s = 1}^{4} \sum_{j=1}^{2L} L_{\alpha, s, j} \rho L^\dag_{\alpha, s, j} - \frac{1}{2} \{ L^\dag_{\alpha, s, j} L_{\alpha, s, j}, \rho \} 
} 
(see SM~\cite{supmat} for a derivation starting from the underlying quantum channels). Here, we take the cluster state $\ket{\psi}_C$ as our initial state but in fact generic mixed initial states evolve to the same steady-state (see SM~\cite{supmat}). 

The resulting dynamics can be intuitively understood from a mean-field analysis, where we ignore inter-site correlations between erasures and defects. In this limit (see SM\cite{supmat} for details): 
\eq{}{\eqsp{
\mathrm{d} n^e/\mathrm{d}t = \ &(\eta -2\gamma n^e)(1-n^e), \\
\mathrm{d} n^d/\mathrm{d}t = \ & \eta(1-2n^d) - 2\gamma n^e(1-n^e).
}}
where $n^d$ ($n^e$) represents the mean-field density of defects (erasures). The steady-state solution is:
\eq{eq:mf-main}{ 
 (n^e,n^d) = \begin{dcases} \left(\eta/2\gamma , \, \eta/4\gamma \right) \qquad &\text{if  }\ \eta <2\gamma,  \\ \left(1, \, 1/2 \right)  \qquad &\text{if  }\ \eta \geq 2\gamma, \end{dcases}
}
which suggests a transition between a phase where both erasures and defects are controlled by the correction protocol ($\eta < 2 \gamma$), and a phase where erasures proliferate ($\eta > 2 \gamma$), at which point the correction protocol fails and defects return to their un-corrected equilibrium density of $1/2$. The existence of a phase in which our correction protocol successfully controls $n^d$ suggests that in this regime, stabilizer defect pairs are effectively confined, and string-order is stabilized in the steady-state.  

We corroborate this mean-field analysis via Monte Carlo simulations on the coupled dynamics of erasures and defects (see SM~\cite{supmat} for details). The resulting steady-state values for the string-order, $n^d$, and $n^e$ as a function of $\eta/\gamma$ are shown in Fig.~\ref{mainfig-1}b. All three quantities simultaneously fall to their respective infinite-temperature values at $\eta_C = 0.6065 \gamma$, indicating a continuous phase transition between a mixed-state SPT phase with non-trivial string-order for $\eta < \eta_C$, and a thermal steady-state with no SPT order for $\eta> \eta_C$.  

Finally, we investigate the nature of this phase transition: the proliferation of erasures is an absorbing state transition of the type discussed in Refs.~\cite{iadecola2022chaos,chen2023steering,khemani2024absorbing,sierant2023absorbing} and can be mapped to directed percolation, for which the critical exponents are known~\cite{hinrichsenNonequilibriumCriticalPhenomena2000}. Our numerical results recover these values (see SM~\cite{supmat}). Using the scaling approach of Ref.~\cite{mendonca2011}, we can determine the critical exponents for $n^d$ and the string-order. Near the phase transition, these are expected to satisfy the scaling relations 
\eq{}{\eqsp{
( 0.5 - n^d(t, L))  \sim & \, t^{- \delta_d } f_d\left ( (\eta - \eta_C)^{\nu_t} t , L^{-z} t \right ) \, , \\  \Omega(t, L)  \sim & \, t^{- \delta_\Omega } f_{\Omega}\left ( (\eta - \eta_C)^{\nu_t} t , L^{-z} t \right ) \, .
}
}
Finite-size scaling indicates that $z = 1.58$ (see SM~\cite{supmat}). The scaling collapse for the remaining exponents $\delta_d, \delta_{\Omega}$, and $\nu_t$ is shown in Fig.~\ref{Fig:Critical}. All critical exponents for $n^d$ match those of the erasures, indicating that defects proliferate in the same directed percolation transition as the erasures. The exponents $\nu_t$ and $z$ associated with the onset of string-order also have values as for directed percolation, while we find $\delta_{\Omega} = 2 \delta_{d}$. This follows by observing that string-order is destroyed by processes that create domain wall pairs across only one of the string's two end-points: thus, with perfect heralding $\Omega_{i,i+l} \sim (1-n^e_{2i})(1-n^e_{2i+2l})$ is essentially a 2-point correlator of the erasures, which for large $l$ approaches $(1-n^e)^2$ (see SM~\cite{supmat} for technical details).  

\emph{Stability.---}We now consider stability of the SPT-ordered steady-state against local perturbations to its dynamics. Since string-order vanishes in the absence of strong symmetry even for a finite-depth local channel~\cite{coser2019class,degroot2022og}, our SPT steady-state is unstable against symmetry-breaking $X$ errors. This is analogous to closed quantum systems, where there are no non-trivial bosonic phases in 1D in the absence of symmetry~\cite{chen2011SPT}. To study stability against strongly symmetric noise, we consider unheralded $Z$ errors which can arise experimentally from imperfect heralding or from unitary dynamics separating defects from the underlying erasures. We set the rate of heralded errors in Eq.~\eqref{Eq:HerErr} to $\eta f_e$, with unheralded errors described by 
the Lindblad operators $L_{f,1,j} = \sqrt{\eta(1-f_e)} Z_j$. The total Pauli error rate is thus $\eta$, with $f_e$ the fraction of heralded errors. 
Fig.~\ref{Fig:unbiased}a shows the string-order at time $t = 2/ \eta_c$ as a function of the heralding fraction $f_e$ and the noise to correction ratio $\eta/ \gamma$. At low noise rates and $f_e \sim 1$, the string-order at this time-scale is greatly enhanced relative to the unheralded model $(f_e = 0)$. Fig.~\ref{Fig:unbiased}b shows that the string-order decays exponentially in time to its steady-state value with a characteristic time-scale $\tau_{\Omega} \sim 1/(1- f_e)$.  
However, for $f_e < 1$ our model does not realize a true steady-state SPT since the steady-state string-order decays exponentially with the string's length (Fig.~\ref{Fig:unbiased}c), with a characteristic length scale $\xi_{\Omega} \sim 1/\sqrt{1-f_e}$. 
In the SM~\cite{supmat}, we show that despite the lack of steady-state string-order, the absorbing state transition persists for $f_e < 1$, and that in the phase where erasures have not proliferated, our correction protocol substantially reduces both $n^d$ and the maximum length of error strings in the steady-state. We also show that signatures of this critical point are detectable from the finite-time behavior of the string-order parameter, which exhibits a crossover from characteristic power-law decay to an exponential decay given by  $\Omega(t)\sim t^{-\delta_\Omega}e^{-4\eta(1-f_e)t}$ (see SM~\cite{supmat}).


\emph{Discussion.---}In this work, we have demonstrated that biased erasure noise permits the stabilization of nontrivial zero-temperature phases as steady-states under local Lindblad dynamics. Crucially, our experimentally motivated noise model can be exploited to effectively confine the errors which otherwise proliferate at any finite temperature or error rate (resulting in a trivial steady-state), thereby enabling local error correction into an SPT phase starting from generic initial mixed states. We have also shown that our protocol significantly enhances the lifetime of string-order in the presence of unheralded errors. Although we have focused on 1D SPT order here, our results extend more broadly to nontrivial ordered states in higher dimensions which lack thermal stability; in particular, heralding enables local error-correction into 2D topologically ordered steady-state phases, which we discuss in Ref.~\cite{chirame2024nonabelian}.


\emph{Acknowledgements: }  The authors are grateful to Tim Hsieh, Sanjay Moudgalya, Ramanjit Sohal, David Stephen, and Jeff Thompson for stimulating discussions. F.J.B. and S.C. are grateful for the support of NSF DMR-1928166 and NSF DMR-2313858. S.G. is supported in part by an Institute for Robust Quantum Simulation (RQS) seed grant. This material is based upon work supported by the Sivian Fund and the Paul Dirac Fund at the Institute for Advanced Study and the U.S. Department of Energy, Office of Science, Office of High Energy Physics under Award Number DE-SC0009988 (A.P.). The authors acknowledge the computational resources provided by the Minnesota Supercomputing Institute (MSI) at the University of Minnesota.


\newpage

\let\oldaddcontentsline\addcontentsline
\renewcommand{\addcontentsline}[3]{}
\bibliography{ref}
\let\addcontentsline\oldaddcontentsline



\onecolumngrid 
\clearpage
\makeatletter 

\begin{center}   
	\textbf{\large Supplementary Material for ``Stable Symmetry-Protected Topological Phases in Systems with Heralded Noise"}\\
	[1em]
	Sanket Chirame$^1$, Fiona J. Burnell$^{1}$, Sarang Gopalakrishnan$^{2}$, and Abhinav Prem$^{3}$ \\[.1cm]
	{\itshape \small ${}^1$School of Physics and Astronomy, University of Minnesota, Minneapolis, Minnesota 55455, USA \\ 
	${}^2$Department of Electrical and Computer Engineering, Princeton University, Princeton, NJ 08544\\
    ${}^3$School of Natural Sciences, Institute for Advanced Study, Princeton, New Jersey 08540, USA}\\
	(Dated: \today)\\[1cm]
\thispagestyle{titlepage} 
\end{center} 	
\setcounter{equation}{0} 
\setcounter{figure}{0} 
\setcounter{table}{0} 
\setcounter{page}{1} 
\setcounter{section}{0} 
\renewcommand{\theequation}{S\arabic{equation}} 
\renewcommand{\thetable}{S\arabic{table}} 
\renewcommand\thefigure{S\arabic{figure}} 
\renewcommand{\theHtable}{Supplement.\thetable} 
\renewcommand{\theHfigure}{Supplement.\thefigure}

\onecolumngrid

\tableofcontents


\section{Review of the 1D \texorpdfstring{$\mathbb{Z}_2 \times \mathbb{Z}_2$}{Z2Z2}  Cluster State SPT}
\label{sec:cluster}

Here, we briefly review the physics of the one-dimensional (1D) $\mbZ_2 \times \mbZ_2$ cluster state SPT~\cite{briegel2001,son2012cluster}. Consider a chain of length $2L$ with an on-site Hilbert spanned by a qubit (or spin-$1/2$) i.e., $\mathcal{H} = \otimes_i^{2L} \mathcal{H}_i$ with $\mathcal{H}_i = \mathbb{C}^2$. The cluster state Hamiltonian is given by
\be
\label{eq:HC}
H = -\sum_{j=1}^{2L} X_{j-1} Z_j X_{j+1} \,,
\ee
where $X, Z$ are standard Pauli operators (with $XZ=-ZX$). Each term of $H$ commutes with all other terms i.e., they are mutually commuting stabilizers and we can straightforwardly analyse its properties. First, note that $H$ has a $\mbZ_2 \times \mbZ_2$ global symmetry, generated by
\be
\mathbb{Z}^a_2 = \prod_{j=1}^{L} Z_{2j} \, ,\quad \mathbb{Z}^b_2 = \prod_{j=1}^{L} Z_{2j-1} \, .
\ee

The ground state of $H$ is simply given by the state $\ket{\psi}$ that is stabilized by the individual terms of the Hamiltonian. On periodic boundary conditions, there is a unique ground state which can be conveniently expressed in terms of a unitary acting on a product state as
\be
\label{eq:psi_C}
\ket{\psi}_{C} = \prod_{j=1}^{2L} CX_{j,j+1} \ket{\uparrow}^{\otimes 2L} \, ,
\ee
where $Z \ket{\uparrow} = \ket{\uparrow}$, and  $CX_{j,j+1} = \frac{1}{2}(1+X_{j}+X_{j+1}-X_j X_{j+1})$ is the entangling gate. To see that the cluster state is the ground state of $H$ in Eq.~\eqref{eq:HC}, note that conjugating the Hamiltonian with the cluster entangler $U = \prod_{j=1}^{2L} CX_{j,j+1}$ returns a paramagnetic Hamiltonian (on periodic boundary conditions): $U H U^\dagger = -\sum_{j=1}^{2L} Z_j$~\cite{briegel2001}. Since $\ket{\uparrow}^{\otimes 2L}$ is the ground-state for the paramagnet, $\ket{\psi}_{C}$ is the ground-state for the cluster Hamiltonian. The structured entanglement of this state is encoded in the following non-local string-order operator, which is given by product of stabilizers over a finite-region of length $l = |i-j|$:
\be
\label{eq:string_order}
\Omega_{ij}= X_{2i} \left(\prod_{k=i}^{j} Z_{2k+1} \right) X_{2j+2} \nonumber
\ee
where
\be
\lim_{|i-j|\to\infty} \braket{\psi_C|\Omega_{ij}|\psi_C} = 1
\ee
More generally, we can consider adding arbitrary local $\mbZ_2 \times \mbZ_2$ symmetric perturbations to the Hamiltonian, and as long as the bulk gap does not close, $ \braket{\psi_C|\Omega_{ij}|\psi_C} = c$, where the constant $0 < c \leq 1$ as $|i-j| \to \infty$~\cite{pollmann2012}. 

Let us now study the system on open boundary conditions. In this case, the cluster entangler is given by $U = \prod_{j=1}^{2L-1} CX_{j,j+1}$, such that 
\be
U H U^\dagger = - \sum_{j=2}^{2L-1} Z_j \, , 
\ee
i.e., this unitary transformation yields a trivial paramagnetic Hamiltonian but the terms at $j=1$ and $j=2L$ are missed. This encodes the fact that the cluster Hamiltonian has a four-fold degenerate ground-state manifold on open boundaries, where this degeneracy corresponds to the anomalous boundary states of the $\mbZ_2 \times \mbZ_2$ SPT state. The missing edge operators (or ``dangling operators") $X_1, Z_1, X_{2L}, Z_{2L}$ in the paramagnet can then be translated back to the original basis by applying the cluster entangler,
\be
\bar{Z}_1 = Z_1 X_2 \, , \quad \bar{X}_1 = X_1 \, , \quad \bar{Z}_2 = X_{2L-1}Z_{2L} \, , \quad \bar{X}_{2L} = X_{2L} \, ,
\ee
where $\bar{Z}_1, \bar{X}_1, \bar{Z}_2,\bar{X}_2$ commute with the Hamiltonian $H$ and satisfy the $\mbZ_2$ algebra $\bar{Z}_{1,2} \bar{X}_{1,2} = - \bar{X}_{1,2} \bar{Z}_{1,2}$. Hence, these are the logical operators for the two logical qubits encoded in the codespace spanned by the degenerate ground-state manifold. It follows from the above discussion that the symmetry is realised projectively on the two edges of the chain, diagnosing the non-trivial SPT order.

\subsection{Stability of SPT order under finite-depth local quantum channels}

We now consider preparing the system in the cluster state $\rho_C = \ket{\psi}_C \bra{\psi}_C$, subject it to local noise, and ask whether the resulting state is in the same mixed state phase of matter. Here, we consider two states $\rho_1, \rho_2$ as being in the same mixed state phase if they are ``two-way" connected i.e., if there exist two symmetric finite-depth local channels $\mathcal{E}_1$ and $\mathcal{E}_2$ such that $\rho_2 = \mathcal{E}_1 (\rho_1)$ and $\rho_1 = \mathcal{E}_2 (\rho_2)$~\cite{coser2019class,sang2023mixed}.

Consider first a symmetry breaking dephasing channel i.e., where the dephasing is given at each site by a Pauli-$X$ operator. This is described by the local error channel $\mE_j: \rho \to \sum_{\alpha_j=0,1} p_{\alpha_j} K_{\alpha_j}^\dagger \rho K_{\alpha_j}$, where $K_{\alpha_j} = X_j^{\alpha_j}$ and $p_{\alpha_j} = 1-p, p$ for $\alpha_j = 0,1$. Here, $p=1/2$ corresponds to maximal dephasing. Thus, the total noise channel on the entire chain is encoded in $\mE = \mE_1 \circ \dots \circ \mE_{2L}$. We can express this as
\be
\mE(\rho) = \sum_{\vec{\alpha}} P(\vec{\alpha}) K^\dagger_{\vec{\alpha}} \rho K_{\vec{\alpha}} \, ,
\ee
where 
\be
P(\vec{\alpha}) = \prod_{j=1}^{2L} p_{\alpha_j} \, , \quad K_{\vec{\alpha}} = \prod_{j=1}^{2L} X_j^{\alpha_j} \, .
\ee
Starting in the pure cluster state, the decohered state is given by $\rho_\mE = \mE(\rho_C)$, such that the expectation value of any operator $\mathcal{O}$ is given by
\be
\braket{\mathcal{O}}_{\rho_\mE} = \sum_{\vec{\alpha}} P(\vec{\alpha}) \bra{\psi}_C  K_{\vec{\alpha}} \mathcal{O} K^\dagger_{\vec{\alpha}} \ket{\psi}_C \, .
\ee
One can then show that the expectation value of the string-order parameter in the decohered state is
\be
\braket{\Omega_{ij}}_{\rho_\mE} \sim (1-2p)^{|i-j|} \braket{\psi_C|\Omega_{ij}|\psi_C} \, ,
\ee
since each $Z$ operator in the bulk of the string-order parameter can anti-commute with the noise operator $X$. In this case, no symmetric quasi-local channel can recover the pure cluster state for any $p \neq 0$~\cite{hseih2023mixed} and the decohered state is in the trivial mixed phase. On the other hand, if we consider a symmetric noise channel i.e., $K_{\alpha_j} = Z_{j}^{\alpha_j}$, then one instead obtains
\be
\braket{\Omega_{ij}}_{\rho_\mE} \sim (1-2p)^2 \braket{\psi_C|\Omega_{ij}|\psi_C} \, ,
\ee
since only the end operators $X$ in the string-order parameter now anti-commute with the noise operator $Z$. Thus, under a single application of a noise channel that respects the strong $\mbZ_2 \times \mbZ_2$ symmetry, the string-order parameter remains finite as long as $p < 1/2$ and the original cluster state is recoverable via a finite-depth local quantum channel~\cite{hseih2023mixed}. However, it should be clear that the string-order parameter will decay exponentially with time under repeated applications of the noise channel even for strongly symmetric noise i.e., $\mE^n(\rho_C) \sim (1-2p)^{2n} \braket{\psi_C|\Omega_{ij}|\psi_C}$. As we discuss in the main text however, one can harness heralded erasures to stabilize a non-zero value of the string-order parameter in the steady state.


\section{Biased Erasure Noise in Rydberg Atom Arrays}
\label{sec:noise}

In this Section, we review the details of the biased-erasure noise model presented in Ref.~\cite{wu2022,sahay2023}, and describe how this leads to the noise channel presented in the main text. Biased erasure noise has recently been discussed in the context of Yb atom qubits, where the $6s6p \ {}^3P_0$ $F=1/2$ atomic levels of neutral ${}^{171}\text{Yb}$ atom are used to encode the computational states $\{|0\rangle ,|1\rangle\}$ of the qubit. These atoms can be arranged in the desired lattice structure using optical tweezers. A recent experimental work~\cite{maHighfidelityGatesMidcircuit2023} has demonstrated high-fidelity gates in this platform. The two-qubit gate error is measured to be $2.0(1)\times 10^{-2}$, where the dominant error process is spontaneous decay from a highly excited Rydberg level (as discussed below). Using fast imaging techniques ($~\sim 20 \mu$s), these errors can be detected and converted into erasures. The single qubit gate operates on a timescale of $2$ ms and can be performed with a total error of $ 1.0(1)\times10^{-3} $. The idling error rate is much smaller due to the long lifetime of the qubit states, which is of the order $\sim 2.96(12)$s. 

In this platform, a two-qubit gate is implemented by selectively coupling the computational state $|1\rangle$ to a highly-excited Rydberg state $|r\rangle$. The enhanced van der Waals interaction between the Rydberg atoms creates a blockade effect that prohibits the simultaneous excitation of nearby atoms to the level $|r\rangle$, and can thus be used to enact two qubit control- gates.  The resulting native 2-qubit gates are control-$Z$ gates, whereas our preparation protocol requires control-$X$ gates, suggesting that Hadamard gates, which convert $Z$ errors to $X$ errors and thus remove the strong symmetry constraint, are necessary for state preparation. We note, however, that proposals for bias-preserving CNOT gates in Rydberg atoms do exist \cite{cong2022}. The gate fidelity is fundamentally limited by the processes where the state $|r\rangle$ decays with probability $2p_e$ to atomic levels that are orthogonal to the computational subspace, creating an erasure error. The location of this error can now be detected by observing fluorescence on the relevant closed cycling transitions; since these transitions happen between levels that are outside the qubit sub-space, errors can be detected without altering the logical state of the qubit. 

When an erasure is detected at a site labeled by $j$, the corresponding qubit is re-initialized to the state $|1\rangle$ since the state $|0 \rangle$ does not couple to $|r\rangle$. The quantum channel acting on the the state of system conditioned on the detection of an erasure is given by $\rho\rightarrow2p_e |1\rangle\langle1|\rho |1\rangle\langle1|$. This noise channel can be further simplified by applying an additional single-qubit Pauli twirl channel 
\eq{}{\mathcal{E}(\rho)\ \rightarrow\ \mathcal{E}_{\text{twirl}}(\rho) = \frac{1}{4} \left[\ \mathcal{E}(\rho) + X \mathcal{E}( X \rho X) X + Y \mathcal{E}(Y \rho Y) Y + Z \mathcal{E}(Z \rho Z) Z\ \right] \,.}
This transforms the original noise channel into a biased stochastic Pauli channel given by
\eq{}{
\rho\rightarrow \frac{p_e}{2} (\rho + Z_j\rho Z_j) \, ,
}
where $j$ is the location of the qubit that experiences the error, and $p_e $ is the probability of an erasure error occurring under a single application of the channel. We introduce a classical variable $n^e_j$ to record the presence $(n^e_j=1)$ or the absence $(n^e_j=0)$ of an erasure at site $j$. The erasure locations thus provide an additional bit of information along with the error syndromes that we will utilize later in our correction protocol. The combined resetting and updates of the erasure record is encoded in the following channel:
\eq{eq:jump}
{\rho \rightarrow \mathcal{E}^e_j(\rho) = \frac{p_e}{2} (Z_j e^-_j \rho e^+_j Z_j + e^-_j \rho e^+_j + Z_j n^e_j \rho n^e_j Z_j + n^e_j \rho n^e_j),}
where $e^-_j = |n^e_j=1\rangle\langle n^e_j=0|$ generates an erasure at an empty site and $e^+_j = |n^e_j=0\rangle\langle n^e_j=1|$ removes the erasure from an occupied site. Here, the first two terms in Eq.~\eqref{eq:jump} describe the action of the Pauli channel on qubits that do not have any erasures before the channel is applied. The latter two terms correspond to qubits that had an erasure before the channel is applied, for which the erasure records are not updated. 

To obtain the error channel in the main text, we must also account for the fact that the continuous monitoring of the erasures leads to non-hermitian evolution on the system even when the erasure is not observed (corresponding to no-jump events). Under the Pauli-twirl approximation, this can be modeled as 
\eq{eq:no-jump}{\rho \rightarrow \mathcal{E}^0_j (\rho)  = \left(1-p_e-\frac{p_e^2}{4}\right)\rho + \frac{p_e^2}{4} Z_j\rho Z_j \ .}
Combining the channels (\ref{eq:jump}) and (\ref{eq:no-jump}), and  assuming a constant rate of decay from the level $|r\rangle$ such that $p_e =\eta \ d t$, we obtain
\eq{}{\rho(t+dt) = \rho(t) - \frac{\eta dt}{2}\left( Z_j e^-_j \rho(t) e^+_j Z_j + e^-_j \rho(t) e^+_j + Z_j n^e_j \rho(t) n^e_j Z_j + n^e_j \rho(t) n^e_j  -2\rho(t)\right) + \frac{(\eta dt)^2}{4}(\rho(t)+Z_j\rho(t)Z_j).}
 
If the time-step $d t$ is sufficiently short, we can neglect terms $\sim O\left({dt ^2}\right)$ and arrive at a continuous time master equation in the Lindblad form, given by
\eq{}{\frac{d\rho}{dt} = \frac{\eta}{2}\left( Z_j e^-_j \rho(t) e^+_j Z_j + e^-_j \rho(t) e^+_j + Z_j n^e_j \rho(t) n^e_j Z_j + n^e_j \rho(t) n^e_j  -2\rho(t)\right).}
The Lindblad jump operators $L_{\eta,k,j}$ in Eq.~(1) from the main text can be read from this expression. The dynamics of the entire system under the noise channel is hence governed by the uncorrelated action of this Lindbladian at each site, encoded by Eq.~(4) in the main text ($\alpha=\eta$ terms):
\eq{}{\frac{d\rho}{dt} = \mathcal{L}_{\text{noise}}(\rho)=\sum_{s=1}^4\sum_{j=1}^{2L} \left( L_{\eta,s,j} \rho L_{\eta,s,j}^\dagger - \frac{1}{2}\{L_{\eta,s,j}^\dagger L_{\eta,s,j},\rho\}\right).}

In summary, the noise process at a particular site $j$ corresponds to a dephasing channel that generates stabilizer defects at adjacent sites $j\pm1$. Concurrently, the location of this event is recorded via the creation of an erasure flag at site $j$.


\section{Details of the Correction protocol}
\label{label:correction}

\begin{figure}
\subfloat{%
  \includegraphics[width=0.6\columnwidth]{{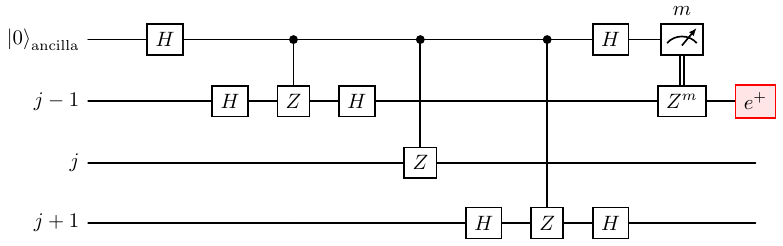}}%
}
\caption{\label{Fig:circuit} The quantum circuit that implements the correction protocol using the native gate-set (Hadamard ($H$) and controlled-Z) of the Rydberg atom platform. The ancilla qubit, initialized in the $|0\rangle_{\mathrm{ancilla}}$ state, is sequentially entangled with system qubits located at $j-1,j$, and $j+1$. A measurement of the ancilla in the computational basis performs a measurement of the stabilizer $S_j=X_{j-1}Z_{j}X_{j+1}$ on the system qubits. The feedback $Z_{j-1}$ is applied conditioned on the measurement outcome $m\in\{0,1\}$. Subsequently, the flag at site $j-1$ is lowered from $1$ to $0$ at the end of the process. This particular circuit shows the application of the jump operator $\frac{1}{2}e^+_{j-1}Z^m_{j-1}(1+(-1)^mS_j)(1-n^e_{j+1})$ when the defect is corrected in the direction of $j-1$. The rest of the correction operators can be designed analogously. Note that while the total measurement+feedback operator obeys the strong symmetry, the component gate $H$ is not strongly symmetric and can hence potentially convert $Z$ errors into $X$ errors. It would be desirable to find ways to develop native gates that obey the symmetry condition and preserve the noise bias. The circuit diagram is drawn using the Quantikz pacakge~\cite{kay2018tutorial}.}
\end{figure}

Here, we provide the motivation and details behind the choice of correction terms in the Lindbladian presented in the main text. The correction protocol is designed under the assumption that the stabilizer syndromes are generated by the perfectly heralded error model ($f_e=1$) (discussed in the previous section). The strong symmetry constraint of $Z$- errors ensures that the stabilizer syndrome $S_j=-1$ centered at site $j$ is generated by an error on either the left neighbor $(Z_{j-1})$ or the right neighbor $(Z_{j+1})$. The error terms in the Lindbladian generate a mixture of defects that are connected via sites hosting erasures. Our goal is to remove these stabilizer defects using a \textit{local} Lindbladian that is \textit{strongly symmetric} under the $\mbZ_2\times\mbZ_2$ symmetry of the cluster state. These two constraints imply that the defects can only be removed in pairs when they are within some local neighborhood of each other. To do this, we will utilize the record of erasures $n^e$ on sites in the neighborhood of the defects to give the defects a biased motion that can potentially stabilize the string-order in the steady-state by effectively confining the stabilizer defects. 

The local moves are determined based on the measurement outcome of $M_j = \{ n^e_{j-1}, S_j, n^e_{j+1} \}$. The measurement outcomes are recorded in terms of local configurations $\textbf{m}_j= \{0/1,\pm1,0/1\}$. The desired target state is the cluster state (no stabilizer defects) with no flags, represented by $\textbf{m}_j = \{0,1,0\}$ for all $j$. The correction protocol proceeds by applying the measurement channel $\mathcal{M}$ at each site with probability $p_c$, followed by a recovery operation $\mathcal{R}$ 
\eq{}{\rho\rightarrow (1-p_c)\rho+p_c\mathcal{R}\circ\mathcal{M}(\rho).}
Since stabilizer defects are created at the ends of erasure strings, a local measurement outcome $\textbf{m}_j= \{1,-1,0\}$ suggests that a nearby partner defect is likely to be on the left side of $j$;  hence we shift the defect to the left using $Z_{j-1}$ and remove the erasure record by applying the transition operator $e_{j-1}^+$. Similarly, the configuration $\textbf{m}_j = \{0,-1,1\}$ that hosts an erasure on the right-side is updated using $Z_{j+1}e^+_{j+1}$. The Kraus operators that implement these operations are given by
\eq{eq:kraus-crc-s}{K_{1,-1,0} = e^+_{j-1}Z_{j-1}n^e_{j-1}\frac{1-S_j}{2}(1-n^e_{j+1})\ , \qquad  
K_{0,-1,1}= e^+_{j+1}Z_{j+1}n^e_{j+1}\frac{1-S_j}{2}(1-n^e_{j-1}).}

The quantum circuit that implements this measurement and feedback operation is shown in Fig.~(\ref{Fig:circuit}). The error model also generates configurations of the form $\{1,1,0\}$ and $\{0,1,1\}$ where the intermediate site between an erasure and a non-erasure does not have a stabilizer defect. Such configurations result from the part of the erasure error channel that acts as $\mathbb{I}$ on the system. We can safely remove such isolated erasure records by applying transition operators $e^+_{j+1}$ or $e^+_{j-1}$. The corresponding Kraus operators are 
\eq{eq:kraus-crc-f}{K_{1,1,0}= e^+_{j-1}n^e_{j-1}\frac{1+S_j}{2}(1-n^e_{j+1}) \ , \qquad  
K_{0,1,1}= e^+_{j+1}n^e_{j+1}\frac{1+S_j}{2}(1-n^e_{j-1}).}

The remaining four measurement outcomes are left invariant without additional feedback. Let us briefly comment on how this choice is consistent with the goal of correcting into the cluster state. If $\textbf{m}_j=\{0,1,0\}$, then this is already in the correct configuration and should be left as is. A configuration where a defect is present without erasures on either side $(\{0, -1, 0\})$ is forbidden by the perfectly heralded ($f_e=1$) noise model and hence this measurement result will not appear unless unheralded defects were present in the initial state. A configuration with $\textbf{m}_j=\{1,1,1\}$ represents a portion of a longer string of erasures that connects defects separated by a distance larger than $3$ sites and so they are not updated by our local protocol:  removing any of these erasures will destroy the continuous string of flags that enables our correction protocol to confine the pair of defects. 

Finally, we can have configurations of the form $\textbf{m}_j = \{1, -1, 1\}$ where erasures are present on both sides of the defect. We choose to not apply any feedback upon detection of such a configuration. We justify this by making two observations: first, note that the erasures of this configuration do not provide any information about the direction of the nearest defect. If we move the defect in either direction with equal probability and remove the erasure along the corresponding direction, it may result in the defect moving in the wrong direction, taking it farther away from its partner. Moreover, if the defect moves in the wrong direction, the removal of the flag in this process breaks the erasure string that links it to its partner defect, hindering our ability to stabilize the string order by confining the defects. 
Secondly, unless $n^e=1$ for all sites, sequences of such configurations will end somewhere with $\{0,\pm 1,1\}$ or $\{1,\pm 1,0\}$ type configurations that terminate at an active site $n^e=0$. These configurations will get corrected by the operations in Eq.~\eqref{eq:kraus-crc-s} or Eq.~\eqref{eq:kraus-crc-f}. Hence, we expect that if the correction is applied frequently enough (such that a finite number of sites have $n^e=0$), then configurations of the form $\{1,-1,1\}$ will eventually be eliminated by our correction model.

The overall correction channel acting at each site is given by
\eq{eq:include-no-feedback}{\rho\rightarrow (1-p_c)\rho + p_c\sum_{\textbf{m}_j} K_{\textbf{m}_j}\ \rho \ K_{\textbf{m}_j}^\dagger,}
where the sum runs over the eight possible measurement outcomes represented by $\textbf{m}_j$. We assume that the correction operation is applied at some fixed rate $\gamma=p_c/dt$ per unit time. Then using the completeness property of the Kraus operators, the above expression can be equally written as
\eq{}{\frac{\rho(t+dt)-\rho(t)}{dt} = \gamma \sum_{\textbf{m}_j}  K_{\textbf{m}_j}\ \rho(t) \ K_{\textbf{m}_j}^\dagger - \frac{1}{2}\{\ K_{\textbf{m}_j}^\dagger K_{\textbf{m}_j},\rho(t)\}.}
In the main text, we study the dynamics of a system initialized in the cluster state, which only generates a classical mixture of stabilizer eigenstates (refer to Sec.~\ref{sec:MC} for further details). Note that if the measurement is not followed by a feedback operation, the eigenstates of the measurement operator remain unchanged. Anticipating this, we can drop the terms corresponding to $K_{0,1,0},K_{0,-1,0},K_{1,1,1}$ and $K_{1,-1,1}$ since they have no effect on the time evolution of the cluster state. Then, after taking the short time limit $dt\rightarrow 0$, we obtain the Lindblad master equation that represents the correction dynamics. The jump operators of the master equation encoded in Eqs.~(2) and ~(3) of the main text can be read off from the Kraus operators, which we reproduce here for the sake of completeness:
\eq{eq:supp-corr1}{L_{\gamma,1,j}=\sqrt{\gamma/4} \ e^+_{j-1}Z_{j-1} (1 - S_j)(1-n^e_{j+1})\ ,\qquad
L_{\gamma,2,j}=\sqrt{\gamma/4}\ e^+_{j+1}Z_{j+1}(1-S_j)(1-n^e_{j-1})}
and,
\eq{eq:supp-corr2}{L_{\gamma,3,j}=\sqrt{\gamma/4}\ e^+_{j-1}(1+S_j)(1-n^e_{j+1})\ ,\qquad L_{\gamma,4,j}=\sqrt{\gamma/4}\ e^+_{j+1}(1+S_j)(1-n^e_{j-1}).}

In Sec.~\ref{sec:MC}, we provide further details on how this quantum master equation effectively reduces to the \emph{classical} population dynamics of the underlying quantum eigenstates for the present model. The inclusion of no-feedback Kraus operators keeps the results of such population dynamics unchanged. However, they provide a mechanism for removing quantum superpositions among the stabilizer eigenstates present in the initial state. In Sec.~\ref{Sec:InitState}, we discuss how this leads to a steady state distribution that is independent of the initial configuration of the system qubits. 


\section{Details of Monte Carlo simulations}
\label{sec:MC}

In this Section, we provide the details of the numerical simulations used to generate the figures in the main text. We work in a basis where the Liouvillian space is spanned by $|\bm{s},\bm{n^e}\rangle\langle \bm{s'},\bm{{n^e}'}|$, where
\eq{}{|\bm{s},\bm{n^e}\rangle:=|s_1,s_2,\ldots ,s_{2L}\rangle\otimes|n^e_1,n^e_2,\ldots,n^e_{2L}\rangle.}
The states $|\bm{s}\rangle$ are stabilizer eigenstates, where $s_j\in\{1,-1\}$ labels the value of stabilizer at site $j$. The states $|\bm{n^e}\rangle$ are erasure occupation number-states at each site, where $n^e_j\in\{0,1\}.$ In this basis, the Lindblad dynamics has the form:

\eq{}{ \label{Eq:DiagonalDynamics}({|\bm{s',{n^e}'}\rangle\langle \bm{s'',{n^e}''}| \ \mathcal{L}\ \big[\sum_{\bm{s, n^e}} \rho_{ (\bm{s,n^e}), (\bm{s, n^e})}|\bm{s,{n^e}}\rangle\langle \bm{s,{n^e}}|\big]})\ \propto \ \delta_{\bm{s',s''}}\delta_{\bm{{n^e}',{n^e}''}},}
where the constant of proportionality can be read off from the explicit expression of $\mathcal{L}$. In other words, in the $|\bm{s,n^e}\rangle$ basis, the diagonal (population) and off-diagonal (coherences) entries of the density matrix are decoupled and evolve independently of each other. In particular, the dynamics does not generate quantum superpositions of stabilizer eigenstates if these are not initially present. 
Thus, if the initial state of the system is a classical mixture of $|\bm{s,n^e}\rangle$, the dynamics is described by the classical rate equation of the populations $\rho_{ (\bm{s,n^e}), (\bm{s, n^e})}$.  We use this insight to model the time evolution using a classical Monte Carlo method to simulate the trajectories of these basis states.  We introduce a double-ket notation to denote the relevant basis states as $|\bm{s,{n^e}}\rangle\langle \bm{s,{n^e}}|:=|\bm{s,n^e}\rangle\rangle$.

Additionally, the Lindbladian obeys the strong $\mbZ_2^a\times\mbZ_2^b$ symmetry. This leads to a further block-diagonal structure wherein the states with different symmetry charges are decoupled from each other. Without loss of generality, we will work in a fixed charge sector defined by $\mbZ_2^a|\bm{s}\rangle=\mbZ_2^b|\bm{s}\rangle=+|\bm{s}\rangle.$ Note that the cluster state (Eq.~\ref{eq:psi_C}) belongs to this subspace.

To further simplify the dynamics, we note that stabilizer defects on odd and even sites are dynamically decoupled.  
The $Z$-error at an odd (even) site generates stabilizer syndromes on the neighboring even (odd) sites. Similarly, the correction protocol from Eq.~\eqref{eq:supp-corr1} and Eq.~\eqref{eq:supp-corr2} couples erasures on the adjacent odd (even) sites with the stabilizer syndromes on the even (odd) sites.  Thus, the state of the system can always be written in a separable form
\eq{}{\eqsp{\rho =\rho_{\text{odd}}\otimes\rho_{\text{even}}= 
&\sum_{\{s_1,s_3,\ldots;n^e_2,n^e_4,\ldots\}} p_{s_1,s_3,\ldots}^{n^e_2,n^e_4,\ldots}\ |s_1,s_3,\ldots;n^e_2,n^e_4,\ldots\rangle\rangle
\bigotimes \\
&\sum_{\{s_2,s_4,\ldots;n^e_1,n^e_3,\ldots\}} p_{s_2,s_4,\ldots}^{n^e_1,n^e_3,\ldots}\ |s_2,s_4,\ldots;n^e_1,n^e_3,\ldots\rangle\rangle.}}
It should be noted that the stabilizer eigenstates $|s_1,s_2,\ldots\rangle$ themselves can not be written as separable states on any spatially disjoint region. This underlying quantum nature of the Monte Carlo trajectories allows us to realize the non-trivial entanglement structure of a string-ordered state.\\

The time evolution is computed using a standard random-sequential update procedure (eg. \cite{hinrichsenNonequilibriumCriticalPhenomena2000}) as follows:\\
Initialize the state in a configuration of the stabilizers and erasure population $|\bm{s_{\text{even}};n^e_{\text{odd}}}\rangle\rangle:=|s_2,s_4,\ldots,n^e_1,n^e_3,\ldots\rangle\rangle$. 
\begin{algorithmic}
\While {$t\leq t_f$}
\For{$i=1,2,\ldots L$}
    \State sample a number $1\leq j\leq L$ from the uniform distribution to choose the site where $\mathcal{L}$ acts 
    \State sample another random number to choose the update rule for the state where $c_0:=1/(\eta+4\gamma)$:
    \State with probability $\ c_0\eta f_e/2$: Apply $Z_{2j-1}e^-_{2j-1}$ (syndrome with erasure)
    \State with probability $\ c_0\eta f_e/2$: Apply $e^-_{2j-1}$ (erasure without syndrome)
    \State with probability $\ c_0\gamma$: Apply $e^+_{2j-1} Z_{2j-1} n^d_{2j}(1-n^e_{2j+1})$   (ref. Eq.~\eqref{eq:supp-corr1})
    \State with probability $\ c_0\gamma$: Apply $(1-n^e_{2j-1})n^d_{2j}\ Z_{2j+1} e^+_{2j+1}$  (ref. Eq.~\eqref{eq:supp-corr1})
    \State with probability $\ c_0\gamma$: Apply $e^+_{2j-1} (1-n^d_{2j})(1-n^e_{2j+1})$   (ref. Eq.~\eqref{eq:supp-corr2})
    \State with probability $\ c_0\gamma$: Apply $(1-n^e_{2j-1}) (1-n^d_{2j}) e^+_{2j+1}$  (ref. Eq.~\eqref{eq:supp-corr2})
    \State with probability $\ c_0\eta (1-f_e)$: Apply $Z_{2j-1}$ (syndrome without erasure)
\EndFor
\State $t\rightarrow t+\Delta t:=t+\frac{1}{\eta+4\gamma}$
\State Record the values of the observables in the state at the end of the Monte Carlo sweep
\EndWhile
\end{algorithmic}


\section{Mean field analysis}
\label{sec:MF}

In this Section, we derive the mean field equations presented in the main text. As discussed in Eq.~\eqref{Eq:DiagonalDynamics}, the diagonal matrix elements of the density matrix in the stabilizer+erasure basis ($|\bm{s,n^e}\rangle$) are decoupled from the off-diagonal elements. The equation for the time evolution of an observable $\hat{O}$ can be obtained as
\eq{}{\frac{d}{dt}\langle O(t) \rangle:=\frac{d}{dt}\tr{(\hat{O}\ \rho(t))} = \tr{\left( \hat{O} \ \mathcal{L}[\rho(t)] \right)}.}
Since the observables considered in this work are diagonal in the $|\bm{s,n^e}\rangle$ basis, it is therefore sufficient to study the dynamics of the diagonal terms of the density matrix. 

\begin{figure}
\subfloat[\label{subfig:mf-a}]{%
  \includegraphics[width=0.5\columnwidth]{{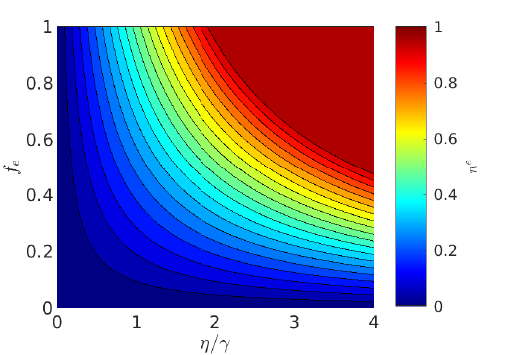}}%
}\hfill
\subfloat[\label{subfig:mf-b}]{%
  \includegraphics[width=0.5\columnwidth]{{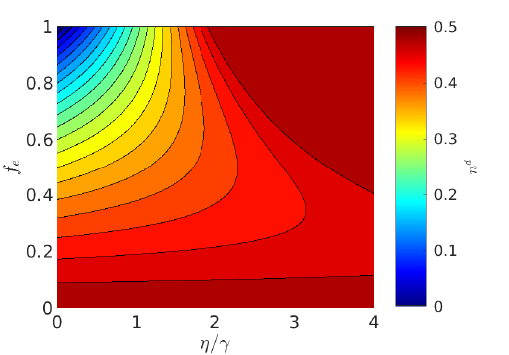}}%
}\hfill
\caption{\label{Fig:MfPhase} Mean field phase diagram: The contour plots of the mean field solution are shown in the $f_e-\eta$ parameter space where $\eta$ is the total noise rate and $f_e$ the fraction of errors that are heralded by erasures. (a) The phase diagram for the density of erasures ($n^e$) shows a continuous transition (see Eq.~\eqref{eq:mf-erasure}) along the curve $\eta f_e=2\gamma$ from an active phase $n^e<1$ to the absorbing phase $n^e=1$. (b) The density of stabilizer defects $n^d$ simultaneously undergoes the transition, with $n^d=1/2$ in the absorbing state (see Eq.~\eqref{eq:mf-nd}).}
\end{figure}

\paragraph{Erasures:} The observable $\hat{n}^e_k$ measures the erasure occupation number at the site labelled by $k$. The exact time evolution equation can be computed using the full Lindbladian, and is given by: 
\eq{}{\frac{\mathrm{d}}{\mathrm{d}t}\langle n^e_k\rangle = \eta f_e (1-\langle n^e_k\rangle) - \gamma \langle n^e_k (1-n^e_{k+2})\rangle-\gamma\langle n^e_k(1-n^e_{k-2})\rangle \, , }
where we take the erasure noise error rate to be $\eta f_e$, with $f_e$ being the heralding fraction. The mean field approximation ignores correlations in the fluctuations at each site. This leads to $\langle n^e_k\rangle\sim n^e$ and $\langle n^e_k n^e_{k'}\rangle\sim {(n^e)}^2$, with the mean-field dynamics given by:
\eq{}{\frac{\mathrm{d} n^e}{\mathrm{d}t} = (\eta f_e-2\gamma n^e)(1-n^e).}
The stable solution in the steady state ($\frac{\mathrm{d} n^e}{\mathrm{d}t}=0$) is
\eq{eq:mf-erasure}{n^e = \begin{cases} \ \frac{\eta f_e}{2\gamma} \qquad &\text{if  }\ \eta f_e<2\gamma,  \\ \ 1 \qquad &\text{if  }\ \eta f_e\geq2\gamma\end{cases}}
Setting $f_e = 1$, we recover the mean-field erasure dynamics in the main text. The mean field phase diagram is shown in Fig.~\ref{subfig:mf-a} and captures the qualitative features of the results obtained using the Monte Carlo simulation (see Fig.~\ref{subfig:fea}). The mean field value of the critical noise rate $\eta_c=2$ is larger than the actual value  $\eta_c=0.6065$; the mean field approximation thus overestimates the ability of correction processes to remove the erasures. In particular, the rate of correction in the mean field approximation is proportional to $\gamma n^e(1-n^e)$, but the actual dynamics removes the erasure only if the adjacent site is unoccupied. Close to the transition point, the system hosts long strings of sites that have erasures and hence the correction protocol is effective only at the end-points of such strings, leading to a far lesser effective removal of erasures. This explains why the mean field over-estimates the threshold noise rate $\eta_c$  at which system enters into the absorbing state with $n^e=1$.

\paragraph{Stabilizer defects:}
The occupation number of stabilizer defects, defined by $n^d_k \equiv \frac{1}{2}(1-S_k)$, evolves according to 
\eq{}{ \eqsp{\frac{\mathrm{d} \eval{n^d_k}}{\mathrm{d}t} = 
\eta(2-f_e)(1-2\eval{n^d_k}) - \gamma \eval{n^d_k \big[n^e_{k-1}(1-n^e_{k+1}) + n^e_{k+1}(1-n^e_{k-1})\big]} \\
+ \gamma \eval{(1-2n^d_k)\big[n^e_{k+1}n^d_{k+2}(1-n^e_{k+3}) + n^e_{k-1}n^d_{k-2}(1-n^e_{k-3})\big]}
}}
We can further simplify this using  translational invariance as 
\eq{}{\frac{\mathrm{d} \eval{n^d_k}}{\mathrm{d}t} = \eta(2-f_e)(1-2\eval{n^d_k}) - 2\gamma \langle n^d_k n^e_{k+1}n^d_{k+2}(1-n^e_{k+3})\rangle - 2\gamma\langle (1-n^e_{k-3})n^d_{k-2}n^e_{k-1}n^d_k\rangle .}

In order to arrive at an approximate mean-field picture, we need to estimate the four-site correlation term resulting from the correction protocol. The correction of defects is most effective in the presence of active erasure sites. In this regime, the density of erasures is low and the probability of consecutive sites with erasures is small. This suggests that if a site has an erasure and no erasure adjacent to it, then such erasure will be accompanied with stabilizer defect pair with $1/2$ probability. In this case, we can approximate the first term proportional to $\gamma$ as $\langle n^d_k n^e_{k+1}n^d_{k+2}(1-n^e_{k+3})\rangle\sim\frac{1}{2}\langle n^e_{k+1}(1-n^e_{k+3})\rangle\sim \frac{1}{2}n^e(1-n^e).$ Similarly, the reflection symmetry implies that the second term is also equal to $n^e(1-n^e)/2$.  The resulting mean-field equation is given by
\eq{}{\frac{\mathrm{d} n^d}{\mathrm{d}t} = \eta(2-f_e)(1-2n^d)-2\gamma n^e(1-n^e).}
Using the steady state value of the erasure density from Eq.~\eqref{eq:mf-erasure}, we obtain the density of stabilizer defects in the steady state as

\eq{eq:mf-nd}{
n^d = \begin{dcases} \ \frac{1-f_e}{2-f_e} + \frac{\eta f_e^2}{4\gamma(2-f_e)} \qquad &\text{if  }\ \eta f_e<2\gamma,  \\ \ 1/2 \qquad &\text{if  }\ \eta f_e\geq2\gamma\end{dcases}
}
Setting $f_e = 1$, we again recover the stabilizer defect dynamics discussed in the main text. Figure \ref{subfig:mf-b} shows the resulting mean-field phase diagram as a function of the overall noise rate and erasure fraction, which qualitatively reproduces the phase diagram obtained from our Monte Carlo results shown in Fig.~\ref{fig:fe-eta-phase}.  


\section{Estimation of critical exponents} \label{sec:exponents}
In this Section, we provide a detailed procedure for estimating the critical exponents and scaling forms near criticality. We assume that when the noise rate is close to its critical value $\eta_c$, the spatial and temporal correlation lengths diverge, leading to scaling behavior of the order parameter~\cite{henkelNonEquilibriumPhaseTransitions2008}. We postulate that under the scaling transformation of the noise parameter $(\eta_c-\eta)\rightarrow \lambda (\eta_c-\eta)$  for $\lambda>0$ (where the correction-rate $\gamma$ is set to 1), the observable expectation value $O$ and corresponding length and time-scales transform as
\eq{eq:scaling-hyp}{O\rightarrow \lambda^{\beta}  O, \qquad x \rightarrow \lambda^{-\nu_x}x,\qquad t\rightarrow \lambda^{-\nu_t}t.
}
This scaling hypothesis can be used to write the observable $O$ as a generalized homogeneous function in the scaling regime. Following Eq.~(\ref{eq:scaling-hyp}), we get 
\eq{}{O(t,\eta_c-\eta,L) = \lambda^{-\beta} O'(\lambda^{-\nu_t}t,\lambda(\eta_c-\eta),\lambda^{-\nu_x}L).}
Since this form is valid for all positive values of $\lambda$, we set $\lambda = t^{1/\nu_t}$ to obtain
\eq{}{O(t,\eta_c-\eta,L) = t^{-\beta/\nu_t} O'(1,t^{1/\nu_t}(\eta_c-\eta),t^{-\nu_x/\nu_t}L).}
It is useful to re-write this expression to allow for different functional forms $O_+ (O_-)$ in the active (absorbing) phase.  We then have:
\eq{eq:scaling}{O_{\pm} (t,\eta_c-\eta,L) = t^{-\delta}\  O_{\pm}(|\eta-\eta_c|^{\nu_t}t,\ L^{-z}t).}
which are of the general scaling form given in the main text.  Here, we have introduced the decay exponent $\delta=\beta/\nu_t$ and the dynamical exponent $z=\nu_t/\nu_x$.

\begin{figure}
\subfloat[\label{subfig:ea}]{%
  \includegraphics[width=0.33\columnwidth]{{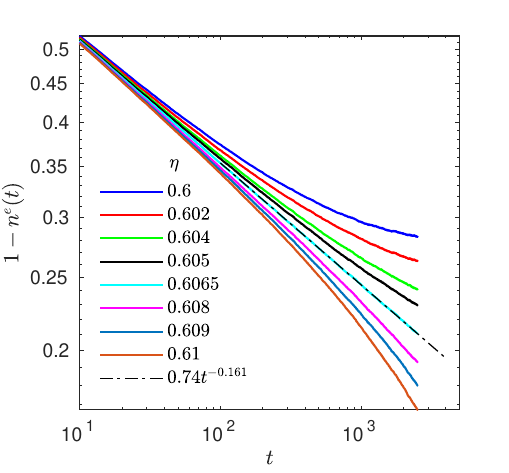}}%
}\hfill
\subfloat[\label{subfig:eb}]{%
  \includegraphics[width=0.33\columnwidth]{{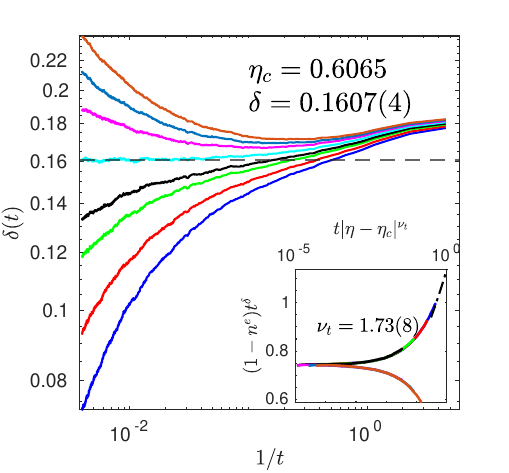}}%
}\hfill
\subfloat[\label{subfig:ec}]{%
  \includegraphics[width=0.33\columnwidth]{{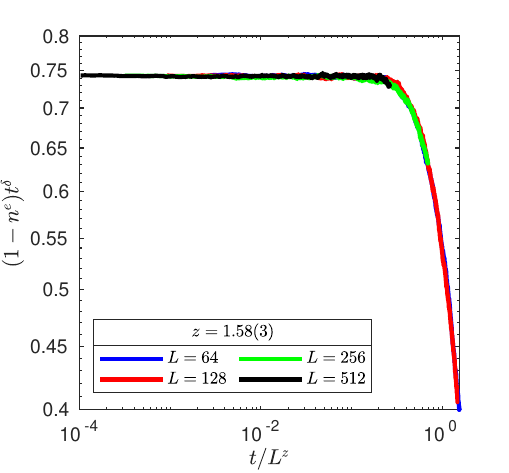}}%
}\hfill
\caption{Critical exponents for the erasure density $n^e$: (a) the density of non erasure sites as a function of time. The data is shown for varying values of the noise rate $\eta$ (with the correction rate $\gamma$ set to 1). The observable at the critical $\eta_c$ decays as a power law $\sim t^{-\delta}$. (b) The instantaneous decay exponent $\delta(t)=\log_{10}{\frac{1-n^e(t)}{1-n^e(10t)}}$ as a function of $1/t$. The $\eta=0.6065$ curve approaches a constant value indicating criticality. The saturation value as $1/t\rightarrow0$ gives the critical exponent $\delta=0.1607(4)$. The inset shows data collapse along the scaling function $O_{\pm}$ when $\nu_t$ is tuned to $1.73(8)$. The system is $L=512$ for the data shown in panels (a) and (b). (c) The finite-size scaling collapse for system sizes $L=64,128,256,512$ when the the dynamical exponent is tuned to $z=1.58(3)$. The results are obtained by averaging over $2\times 10^4$ independent Monte Carlo realizations initialized in the $S_j=1,n^e_j=0, \forall j$ state.}\label{fig:ne_exponents}
\end{figure}

We will follow the procedure provided in \cite{hinrichsenNonequilibriumCriticalPhenomena2000} to  estimate the exponents $\delta$, $\nu_t$, and $z$ from the Monte Carlo data. The exponent for the order parameter ($\beta$) and spatial correlation length ($\nu_x$) can then be obtained using the scaling relations mentioned here.
Specifically, we proceed as follows:
\begin{enumerate}
    \item First, we estimate the instantaneous power-law decay exponent by plotting $\delta(t):= \frac{1}{\log(b)}\log\frac{O(t)}{O(bt)}$ as a function of time for a range of values of $\eta$. The order parameter at the critical point $(\eta=\eta_c)$ goes to zero as $t^{-\delta}$ so $\lim_{t\rightarrow\infty}\delta(t,\eta=\eta_c)=\delta$. In the absorbing phase $(\eta>\eta_c)$ we get $\delta(t)\rightarrow\infty$ and in the active phase $(\eta<\eta_c)$ $\delta(t)\rightarrow0$. This provides the estimate of critical point $\eta_c$ and the exponent $\delta$.  Fig. \ref{subfig:eb} illustrates how this method is used to identify $\eta_c= 0.6065$ using the scaling of the density of erasure flags.  
    
    \item Having found $\delta$, we next plot $O(t) t^\delta$ as a function of $t|\eta - \eta_c|^{\nu_t}$.  The value of $\nu_t$ is obtained by tuning this parameter to obtain  data collapse along the scaling functions $O_{\pm}$. An example is shown in the inset of Fig. \ref{subfig:eb}.
    
    \item The remaining exponent $z$ can be estimated using finite size scaling of the critical curve ($\eta=\eta_c$). To do this, we plot $O(t,\eta=\eta_c,L)t^\delta$ as a function of $t/L^z$ and tune $z$ to obtain the scaling collapse.  An example is shown in Fig. \ref{subfig:ec}.
\end{enumerate}

\subsection{Number of erasures}
When the noise rate is above the critical value, the system enters an absorbing state where all sites have erasures $n^e_j=1$. When $\eta<\eta_c$, the steady state supports finite number of sites that don't have erasure flags. We track this transition using the density of non-erasure sites as
\eq{}{1-n^e(t):=\tr{(\rho(t)  \frac{1}{L}\sum_{j=1}^{L} (1-n^e_{2j-1})} = \frac{1}{L}\sum_{j=1}^{L}(1-\tr{(\rho(t) n^e_{2j-1})}).}
The time evolution of the reduced density matrix of the erasures ($\rho^e = \text{Tr}_{\text{stabilizer}}\rho$) maps on to the classical contact process~\cite{harris1974contact,hinrichsenNonequilibriumCriticalPhenomena2000}. The Lindbladian for this process can be written as 
\eq{eq:contact-L}{\eqsp{\mathcal{L}_{\text{contact}} (\rho^e)= \eta &\sum_j \left(e_j^-\rho^e e_j^+ + n^e_j\rho^e n^e_j - \rho^e\right) 
+\gamma \sum_j\left(e^+_{j-1}(1-n^e_{j+1})\rho^e(1-n^e_{j+1})e^-_{j-1} - \frac{1}{2}\{n^e_{j-1}(1-n^e_{j+1}),\rho^e\}\right)\\
&+\gamma \sum_j\left( e^+_{j+1}(1-n^e_{j-1})\rho^e(1-n^e_{j-1})e^-_{j+1} - \frac{1}{2}\{n^e_{j+1}(1-n^e_{j-1}),\rho^e\}\right).}}
In Fig.~(\ref{fig:ne_exponents}), we show our numerical estimates of the critical exponents, which match well with the values reported previously in the literature~\cite{henkelNonEquilibriumPhaseTransitions2008}.

\begin{figure}
\subfloat[\label{subfig:a}]{%
  \includegraphics[width=0.45\columnwidth]{{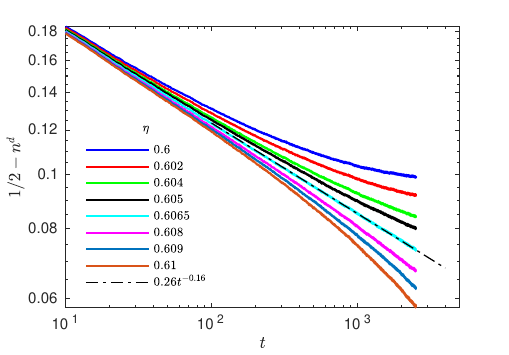}}%
}\hfill
\subfloat[\label{subfig:b}]{%
  \includegraphics[width=0.45\columnwidth]{{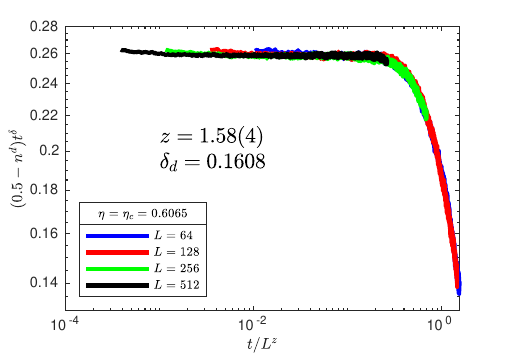}}%
}\hfill
\caption{Critical exponents for the stabilizers defects: (a) The deviation of the density of stabilizer defects $n^d$ from its value in the absorbing state as a function of time. The data is presented for $L=512$ and varying noise rates close to the critical point. The critical curve shows a power law decay $\sim t^{-\delta_{d}}$. The estimation of exponents $\delta_d$ and $\nu_t$ is shown in Fig.~2 in the main text. (b) The dynamical exponent $z=1.58(4)$ is estimated by observing the data collapse of the critical curve $\eta=\eta_c$ for varying system sizes. The data is obtained by averaging over $2\times 10^4$ independent Monte Carlo runs initialized in the perfectly ordered state $S_j=1,n^e_j=0$ for all sites $j$.}\label{fig:nd_exponents}
\end{figure}

\subsection{Density of stabilizer defects}
If $S_j=-1$, then we term that as a stabilizer defect at site $j$. The density of these defects on the even sublattice is tracked using
\eq{}{n^d(t):=\tr{\left(\rho(t) \frac{1}{L}\sum_{j=1}^L n^d_{2j}\right)}=\frac{1}{L}\sum_{j=1}^L\tr{\left(\rho(t)\frac{1-S_{2j}}{2}\right)}.}
When the system is in the absorbing phase, the correction part of the Lindbladian leaves this state invariant and becomes inactive. In this regime, noise stabilizes a maximally mixed state on the system qubits which leads to $n^d=1/2$. In the active phase ($\eta<\eta_c$), the number of stabilizer defects is limited by the number of erasures. The mean field calculation in Sec.~\ref{sec:MF} shows that $\frac{1}{2}-n^d \propto (1-n^e)$. We expect that this functional relation between these \emph{local} observables is valid beyond mean field. This suggests that $\frac{1}{2}-n^d$ will approach zero with the same exponents as the erasures. We numerically confirm this intuitive picture in Fig.~(\ref{fig:nd_exponents}) and Fig.~2 in the main text.

\subsection{String-order}
The SPT order can be diagnosed using the string-order parameter (Eq.~\ref{eq:string_order}) which can be equivalently written as a product of the stabilizers 
\eq{}{\Omega_{i,i+l-1}(t)=\text{Tr}  \left( \rho(t) \prod_{k=i}^{i+l-1}S_{2k} \right ) .}
Thus the string-order over length $l$ is equivalent to the parity of the number of stabilizer defects in the region spanned by this string. Now, let us use this definition to estimate the string-order in terms of the density of erasures $n^e$. In a translationally invariant system, $\Omega_{i,i+l}$ should depend only on the separation $l$ between the string's endpoints. In order to reduce the fluctuations of the string-order while averaging over independent Monte Carlo runs, we therefore use a spatially averaged observable:
\eq{eq:avg-sto}{\overline{\Omega_l}(t) =  \frac{1}{L}\sum_{i=1}^{L}\Omega_{i,i+l-1}(t).}

\begin{figure}
\subfloat[\label{subfig:sto-a}]{%
  \includegraphics[width=0.33\columnwidth]{{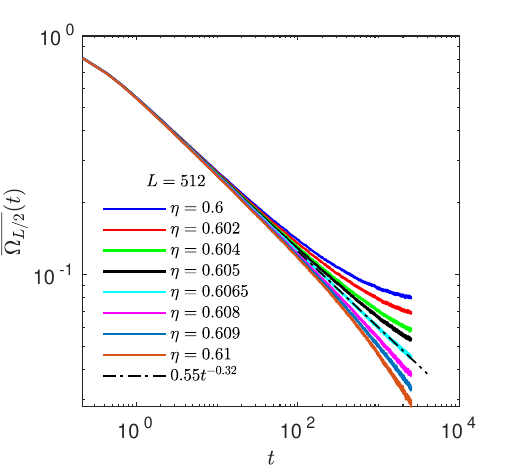}}%
}\hfill
\subfloat[\label{subfig:sto-b}]{%
  \includegraphics[width=0.33\columnwidth]{{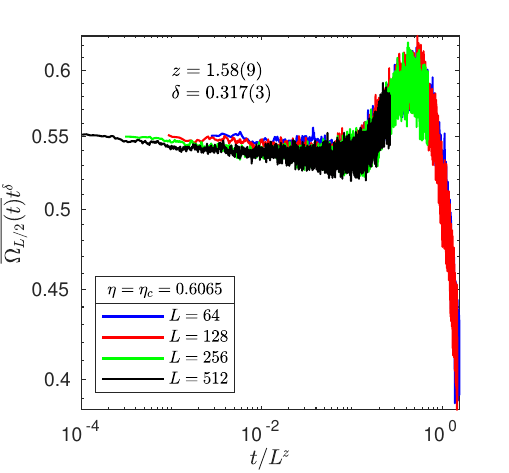}}%
}\hfill
\subfloat[\label{subfig:sto-c}]{%
  \includegraphics[width=0.33\columnwidth]{{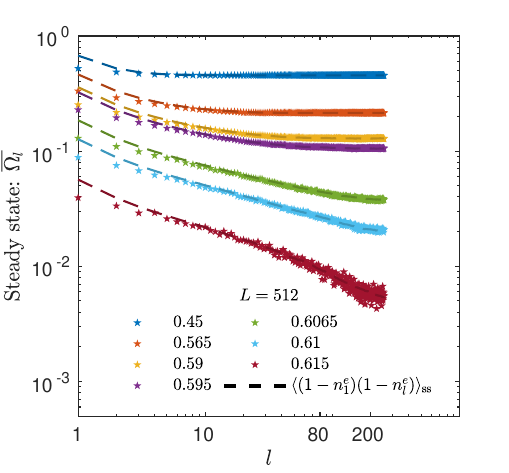}}%
}\hfill
\caption{String order critical exponents: (a) the average value of the string order for length $L/2$ (see Eq.~\eqref{eq:avg-sto}) is plotted as a function of time for noise rates close to critical point. The string order at the critical noise rate decays with a power law $t^{-0.32}$, consistent with the exponent $\delta_\Omega$ obtained in the main text. The data is plotted for $L=512$ sites per sublattice and $4\times 10^4$ independent Monte Carlo runs. (b) The finite size collapse at the critical noise-rate $\eta_c$ is observed when the dynamical exponent is tuned to $z=1.58(9)$. The data is obtained by averaging $2\times 10^4$ Monte Carlo realizations. (c) The steady state value of the string order $\Omega_{l}$ is plotted as a function of the length of the string $l$ for an $L=512$ sized system. The data for a wide range of noise rates matches with the steady state value of the two point correlation function of erasures defined by $\langle(1-n^e_1)(1-n^e_l)\rangle$ (shown using dashed lines of the same colors). The steady state values are obtained by simulating $5\times 10^3$ realizations upto $2L^{1.5}$ MC sweeps.}
\end{figure}

We now explain the relationship between this order parameter and the density of erasure defects, which leads to the relationship $\delta_{\Omega} \sim0.32 =2 \delta_{n^e}=2\delta_{d}$ noted in the main text. The perfect heralding of the stabilizer defects ensures that each defect has an erasure on at least one of its adjacent sites. So the defects are present only within the region that has erasure flags. Additionally, the strong symmetry of the Lindbladian ensures that the defects are created in pairs. Hence the number of defects within these erasure regions is always even. We can now use these two key ideas to estimate the string order in terms of the density of erasures.

Let us consider the string order of a region $\mathcal{D}$ which has endpoints at $i_L$ and $i_R$. If the value of erasures at both endpoints is zero, then we are guaranteed to have an even number of defects inside the region $\mathcal{D}$. All such configurations will result in perfect string order, equal to 1. On the other hand, if there is an erasure at one or both endpoints, then there are two further possibilities. First, the boundary erasure can have stabilizer defects on either side, which results in only one member of the defect-pair being inside the region $\mathcal{D}$. Such configurations will have string order equal to $-1$. Alternatively, such an erasure could have been the result of the $\mathbb{I}$ operation in the noise channel, which creates erasures without creating the syndrome. In this case, the number of defects inside the region $\mathcal{D}$ remains even, resulting in the $+1$ string order. Given the presence of an erasure at the endpoint, both of these outcomes are equally likely such that their effect averages out to zero and they hence do not contribute to the overall string order expectation value.\\
The string order of the region $\mathcal{D}$ can now be estimated as the probability of the absence of erasures at both of its endpoints, which leads to 
\eq{}{\Omega_{i,i+l-1} \ \sim\ \langle (1-n^e_{2i-1})(1-n^e_{2i+2l-1})\rangle.}
The string order is thus equivalent to the two-point correlation function of the order parameter that tracks the erasure transition. In Fig.~\ref{subfig:sto-c}, we numerically verify this relation. When the endpoints of the string are far enough (such that $l>>\xi_x$), we obtain $\Omega\sim(1-n^e)^2$, leading to the observed relationship
\eq{}{\delta_{\Omega}=2\delta_{n^e}.}


\section{Initial state dependence}
\label{Sec:InitState}
In this Section, we further illustrate the stable nature of the string order irrespective of the initial state of the system. In particular, we show that all initial states with zero flags at each site and an arbitrary mixture of stabilizer defects evolve to the same steady state distribution.

The numerical results presented in the main text are obtained using the Monte Carlo simulation of a system initialized in the perfectly ordered state
\eq{eq:cluster-init}{\rho(t=0) = \bigotimes_{j=1}^{2L}|s_{j}=1\rangle\langle s_{j}=1| \bigotimes_{k=1}^{2L}|n^e_{k}=0\rangle\langle n^e_{k}=0|.}
We now wish to argue that the late-time results remain unchanged even if the system is not initialized in the perfect cluster state, which requires performing entangling gates on a product state (see Sec.~\ref{sec:cluster}). These two-qubit gates can themselves lead to errors that create stabilizer defects in the initial state. Hence, it is desirable to ensure that the late time string-order (or absence thereof) is stable against these initial perturbations. Following the arguments in Sec.~\ref{sec:MC}, we work in a fixed symmetry-charge sector $\mbZ_2^a|\bm{s}\rangle=\mbZ_2^b|\bm{s}\rangle=+|\bm{s}\rangle.$

The formal solution of the dynamics generated by the Lindbladian $\mathcal{L}$ is given by 
\eq{eq:exp-L}{\rho(t)=e^{\mathcal{L}t}\rho_0=\sum_{\alpha=0}^{\text{dim}(\mathcal{L})-1}\ e^{\lambda_\alpha t}\ \hat{r}_\alpha \ \tr{\left(\hat{l}^\dagger_\alpha \rho_0\right)}\frac{1}{\tr{\left(\hat{l}_\alpha^\dagger \hat{r}_\alpha \right)}} ,}
where $\rho_0$ is the initial state at $t=0$. The operators $\hat{r}_\alpha$ ($\hat{l}_\alpha$) are the right (left) eigen-operators of the Lindbladian with eigenvalues $\lambda_\alpha$ (see Ref.~\cite{manzano2020} for a brief review). The real parts of all eigenvalues are strictly non-positive. The steady state of the system corresponds to the right eigen-operators with zero eigenvalue. On the other hand, the left eigen-operators with zero eigenvalue capture the information about the initial state that is retained in the steady state. A system going through an absorbing phase transition can have multiple steady states (i.e. eigenmodes with $\lambda=0$) in the active phase ($\eta/\gamma<\eta_c$). This degeneracy implies that the steady state distribution can in-principle depend on the initial conditions in terms of $c_\alpha:=\tr{\left(\hat{l}^\dagger_\alpha \rho_0\right)}$. We now analyze the dependence of coefficients $c_\alpha$ on the initial state for the Lindblad model discussed in the main text.

\begin{figure}
\subfloat[\label{subfig:mix-a}]{%
  \includegraphics[width=0.33\columnwidth]{{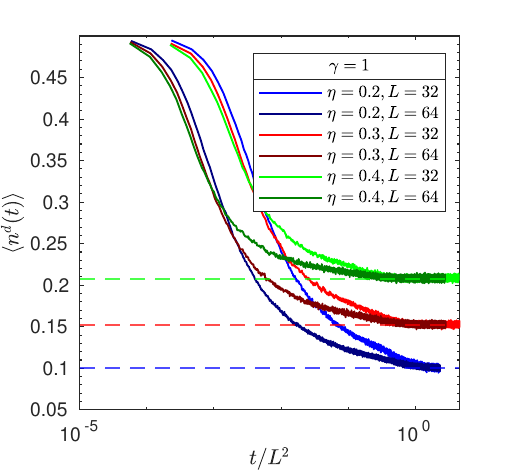}}}\hfill
\subfloat[\label{subfig:mix-b}]{%
  \includegraphics[width=0.33\columnwidth]{{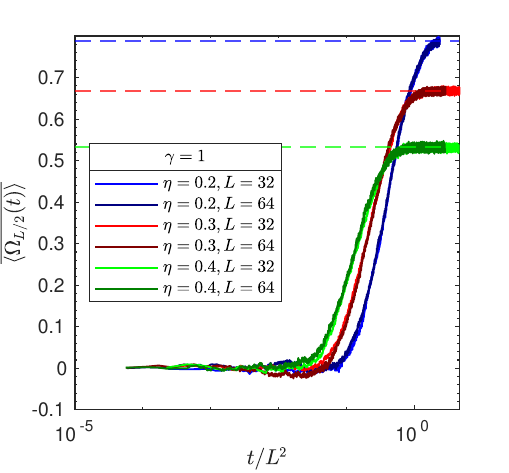}}}\hfill
\subfloat[\label{subfig:mix-c}]{%
  \includegraphics[width=0.33\columnwidth]{{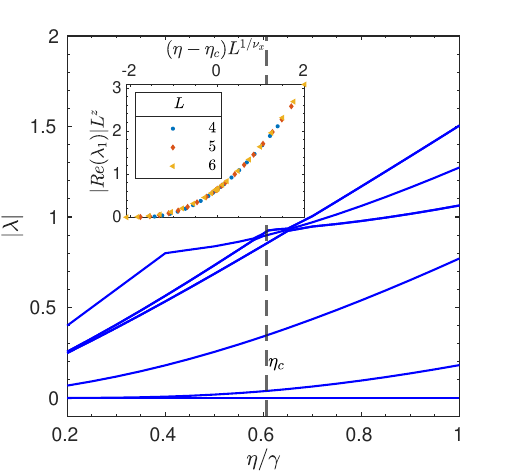}}}\hfill
\caption{Initial state dependence: (a)-(b)The numerical results are obtained by averaging over independent Monte Carlo trajectories initialized in random configuration of stabilizer defects (with even parity) and no erasure flags. The average over these independent samples simulates the dynamics of system initialized in a mixed state given by $\rho=\frac{1}{2}(I+\mbZ_2^a)\otimes_{k=1}^{L}|n^e_{2k-1}=0\rangle\langle n^e_{2k-1}=0|$ at time $t=0.$ The data points are obtained by averaging over $2\times 10^3 - 4\times 10^3$ samples. (a) The density of stabilizer defects and (b) the half-chain string order are plotted as a function of rescaled time $t/L^2$ for varying values of the noise rate $\eta$ and sites per sublattice $L$. The observables saturate to a value identical to the one obtained by initializing the system in the perfectly ordered cluster state (shown using dashed lines). (c) The eigenvalue spectrum of the Lindbladian obtained using the exact diagonalization of an $L=6$ sized system. Here, we include the jump operators that don't have associated feedback operations to obtain the full Lindbladian (see Eq.~\eqref{eq:include-no-feedback}). The gap between the eigenmodes with the largest (i.e. smallest in absolute value) real part closes at the critical point. The inset shows the finite-size scaling collapse of this gap. The exponents used for tuning are obtained using the Monte Carlo simulation.}\label{fig:mixed-init}
\end{figure}

A careful look at the correction terms given in Eq.~\eqref{eq:kraus-crc-s} and~\eqref{eq:kraus-crc-f} shows that the movement of erasures is identical, regardless of the presence or absence of a stabilizer defect on the intermediate site. Similarly, the noise channel creates an erasure with a rate that is independent of the state of stabilizers on neighboring sites. This suggests that the reduced density matrix of the erasures evolves according to 
\eq{}{\frac{d}{dt}\tr_s{\rho} = \tr_s{(\mathcal{L}(\rho))} = \mathcal{L}_e(\tr_s{\rho}),}
where $\tr_s$ represents the trace over the system qubits. The Lindbladian $\mathcal{L}_e$ is nothing but the classical contact process described in Eq.~\eqref{eq:contact-L}, that acts only on the erasure sites. We now use the well-known steady-state structure of the contact process~\cite{hinrichsenNonequilibriumCriticalPhenomena2000} along with stabilizer dynamics to infer the steady state distribution of the full model. 

The absorbing state, where all erasures are occupied ($n^e_j=1,\forall j$), is a steady state of the contact process. In this case, the noise channel acts as dephasing in the $Z$ basis of the qubits. In contrast, the correction channel performs measurements in the stabilizer basis without any feedback. With these non-commuting projection operations on the qubit, we expect that the corresponding steady state of the qubits will be the maximally mixed state within the given symmetry sector. Consequently, the only information about the initial state retained at late times is its global $\mbZ_2^a\times\mbZ_2^b$ charge. The corresponding right and left eigen-operators are given by
\eq{eq:zero-0}{\hat{r}_0 = \frac{1}{4}(1+\mbZ_2^a)(1+\mbZ_2^b)\ \bigotimes_{k=1}^{2L}|n^e_{k}=1\rangle\langle n^e_{k}=1|, \qquad \hat{l}_0 = \frac{1}{4}(1+\mbZ_2^a)(1+\mbZ_2^b)\bigotimes \mathbb{I}. }
When $\eta/\gamma>\eta_c$, this is the unique steady state and all initial conditions evolve to it as $t\rightarrow \infty$. For $\eta/\gamma<\eta_c$, however, there exists a second pair of eigen-operators with zero eigenvalue. This takes the form
\eq{eq:zero-1}{\hat{r}_1 = \tilde{\rho}^R ,  \qquad \hat{l}_1 = \frac{1}{4}(1+\mbZ_2^a)(1+\mbZ_2^b) \bigotimes \hat{\tau}_L^e \, ,}
and describes the active phase of the contact process. The operator $\hat{\tau}^e_L$ acting on the erasure sites is the left eigen-operator of the contact process $\mathcal{L}_e$. We numerically confirm that these two are the only steady states of this model using exact diagonalization of the full Lindbladian. The slowest decay modes close to zero-eigenvalue are shown in Fig.~(\ref{subfig:mix-c})). The $\eta/\gamma<\eta_c$ parameter regime has two eigenvalues ($\lambda_0,\lambda_1$) that take value equal to zero. At the critical point, $\lambda_1$ becomes finite and the system supports a unique steady state for larger values of $\eta$. The scaling collapse in the inset suggests that the gap at $\eta/\gamma=\eta_c$ closes as $L^{-z}$ with increasing system size.

Combining Eqs.~\eqref{eq:zero-0} and~\eqref{eq:zero-1} with the time evolution of the initial state $\rho_0$ (see Eq.~\eqref{eq:exp-L}), the state at late times depends on 
\eq{}{c_0 = \tr{\left(\hat{l}^\dagger_0 \rho_0\right)} = 1,\qquad 
c_1= \tr{\left(\hat{l}^\dagger_1 \rho_0\right)} \propto \tr{\left( \tau^e_L \rho_0\right) \, .}
}
Since $\tau^e_L$ only acts on the erasure flags, the late time state has no information about the stabilizer defects present in the initial state (apart from the global symmetry charge of $\mbZ_2^a\times\mbZ_2^b$). We numerically confirm this in Fig.~\ref{fig:mixed-init} by showing that the system initialized in a maximally mixed configuration of even parity stabilizer defects \emph{and} zero erasure occupations on all sites i.e.
\eq{}{\rho_0 = \frac{1}{4}(1+\mbZ_2^a)(1+\mbZ_2^b)\bigotimes_{k=1}^{2L}|n^e_{k}=0\rangle\langle n^e_{k}=0|}
evolves to a stationary distribution identical to the one obtained using the perfectly ordered initial state in Eq.~\eqref{eq:cluster-init}.


\section{Non-erasure errors}
\label{sec:non-erasure}

Here, we provide further details regarding the effect of non-erasure errors which are generically expected to be present in neutral atom experiments, albeit at a rate that is significantly lower than that of the heralded errors. 

Due to the strong symmetry constraint, we consider biased non-erasure noise that acts as $Z$ errors on the system. This can arise, for example, when all errors are biased erasure errors but the locations of these errors are not detected; hence, there are no erasure flags corresponding to these defects. 
These non-erasure errors contribute to the time evolution of the density matrix via:
\eq{}{
\delta \left( \frac{d}{dt}\rho \right)  = \eta(1-f_e)\sum_j ( \ Z_j\rho Z_j-\rho \ ),}
where $f_e$ is the fraction of non-erasure errors.

\begin{figure}
\subfloat[\label{subfig:fea}]{%
  \includegraphics[width=0.33\columnwidth]{{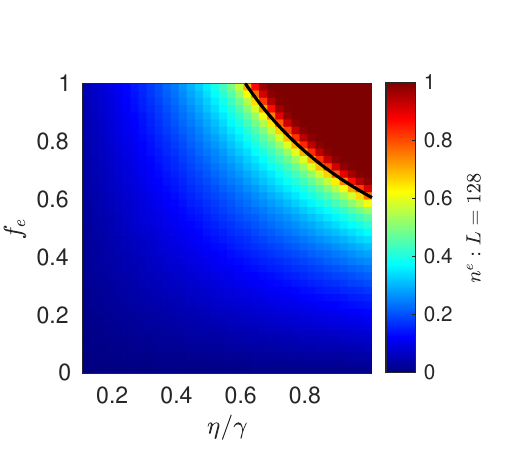}}%
}\hfill
\subfloat[\label{subfig:feb}]{%
  \includegraphics[width=0.33\columnwidth]{{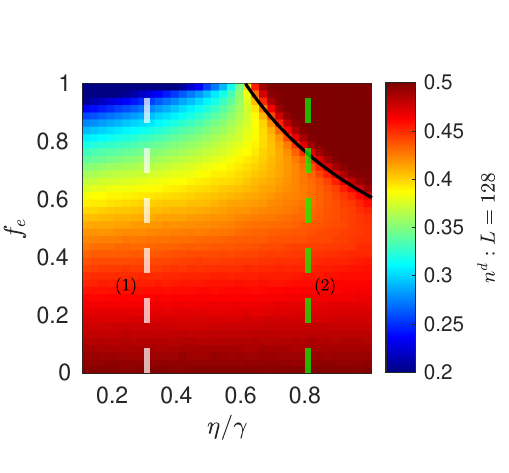}}%
}\hfill
\subfloat[\label{subfig:fec}]{%
  \includegraphics[width=0.33\columnwidth]{{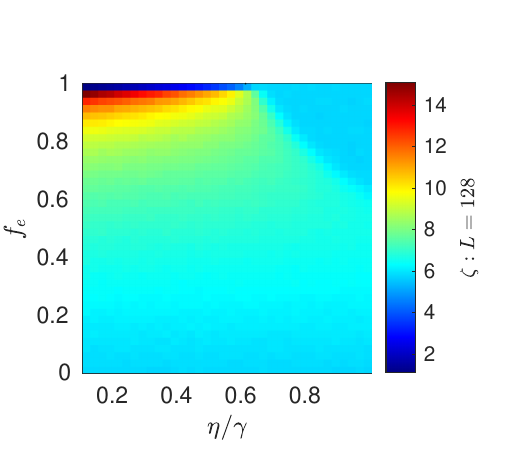}}%
}\hfill
\caption{$L=128$ phase diagram: the color plot in the parameter space defined by the erasure fraction ($f_e$) and the total noise-rate ($\eta/\gamma$) is generated using Monte Carlo simulations. The steady state values are obtained by averaging the data after $2L^{1.5}$ MC sweeps over $5\times 10^3$ independent trajectories. (a) The density of erasures only depends on the rate of the erasure noise $\eta f_e$. The black line shows the corresponding transition curve $f_e=\eta_c/\eta=0.6065/\eta$. (b) The density of stabilizer defects $n^d$. The black curve shows the location of the erasure transition. The white and green dashed lines denote the cuts discussed in the text. (c) The average value of the largest distance between the defects $\zeta$ (see Eq.~\eqref{eq:zeta-mwpm}).}\label{fig:fe-eta-phase}
\end{figure}

The addition of unheralded error processes does not change the dynamics of the erasure sites $n^e$ since these processes only act on the system qubits. Therefore, the absorbing state transition of the erasure sites persists for $f_e<1$. The location of the transition is controlled by the competition between the rate of heralded noise $\eta f_e$ and the correction rate $\gamma$. This means that erasure sites enter the absorbing phase exactly at $\eta f_e/\gamma = \eta_c=0.6065$. We numerically verify this in Fig.~\ref{subfig:fea} using Monte Carlo simulations. 

On the other hand, unheralded error processes do generate stabilizer defects in the system. When this is the only noise process (i.e. $f_e=0$), the erasures remain unoccupied, rendering the correction ineffective. Hence, the steady state is given by the maximally mixed state, with an uncorrelated defect density equal to $1/2$ on each site. Increasing the fraction of erasure noise to $f_e>0$ activates the correction protocol, since heralded errors generate defects flagged by the erasures, which in the active phase can be eliminated by our correction protocol. Thus, increasing the fraction $f_e$ of heralded errors (while keeping the overall noise rate $\eta$ fixed) reduces the steady state density of the defects (see the white dashed line in Fig.~\ref{subfig:feb}), provided that the erasures remain in the active phase.

If the total noise rate, $\eta$ is less than $\eta_c$, it follows that increasing $f_e$ always decreases the density of stabilizer defects.  However, if  $ \eta > \eta_c$, increasing  $f_e$ beyond $\eta_c / \eta$ takes the heralded noise rate beyond its critical value (see the green dashed line in Fig.~\ref{subfig:feb}). In this regime, the erasures proliferate in the steady state ($n^e_j=1\forall j$) and the correction protocol becomes inert. Consequently, the stabilizer defects do not get confined and the steady state is again given by the maximal mixture with $n^d=1/2$.

We find that whether or not heralding decreases the density of stabilizer errors, the steady state string order vanishes for any finite rate of unheralded noise. In Fig.~\ref{subfig:fec}, we show that this occurs because while the correction protocol suppresses the density of stabilizer defects, in the presence of unheralded noise it cannot control the maximum distance between pairs of stabilizer defects in the steady state. Letting $\zeta_{\bm{s}}$ be the largest distance between defects in a particular stabilizer state $|\bm{s}\rangle$, the average of this maximal length in a mixed state $\rho$ can be defined as 
\eq{eq:zeta-mwpm}{\zeta = \sum_{\bm{s}}\rho_{\bm{s}} \zeta_{\bm{s}}.}
This length scale can be interpreted as follows: suppose we want to error correct a state $\rho$ into the cluster state using a single application of a symmetric recovery channel. Then $\zeta$ would correspond to the weight of the largest-weight Pauli-$Z$ operator averaged over syndrome measurement outcomes. A large value of $\zeta$ implies that there are some defect pairs that must be annihilated to achieve this correction, but are far apart from each other. In the presence of a small rate of non-heralded error $f_e\sim 1$ in the active phase ($\eta<\eta_c$), we observe that $\zeta$ can take on relatively large values (see Fig.~\ref{subfig:fec}) even in regions where the overall density of defects is low. This suggests that the string order is destroyed in the steady state due to a small number of defects that are far away from each other. This large separation makes it equally likely to have even or odd numbers of defects in a given extended spatial region, thereby destroying the string order. 

A notable feature of Fig.~\ref{subfig:fec} is that $\zeta$ actually decreases with $f_e$ throughout the active phase. This occurs because as the density of defects increases, $\zeta$ becomes limited by the defect density (since defects that collide annihilate). Thus at large defect densities, we should view the the string order as being destroyed due to the high density of uncorrelated defects, which again makes it equally likely that an even or odd number of defects are contained in any given spatial region that is large compared to the inter-defect spacing.

\subsection{Dynamics of Unheralded Defects} 

To understand these numerical results, we will obtain an effective mean-field like theory from the dynamics of our unheralded defects.  
Let us define $\hat{h}_k = (1-n^e_{k-1})n^d_{k}(1-n^e_{k+1})$ to be an unheralded defect, meaning that it does not have a flags on either side. The density of these defects evolves as 
\eq{}{\frac{d}{dt} \langle h_k\rangle = \frac{d h_k}{dt}\bigg|_{\text{heralded noise}} + \frac{d h_k}{dt}\bigg|_{\text{unheralded noise}} + \frac{d h_k}{dt}\bigg|_{\text{correction}}}
where $\langle .\rangle$ represents the expectation value of the observable in the time evolved mixed state. We now discuss the rate of change of the density of unheralded defects $h_k$ that results from each of these components of our channel.

\begin{figure}
\subfloat{%
  \includegraphics[width=0.95\columnwidth]{{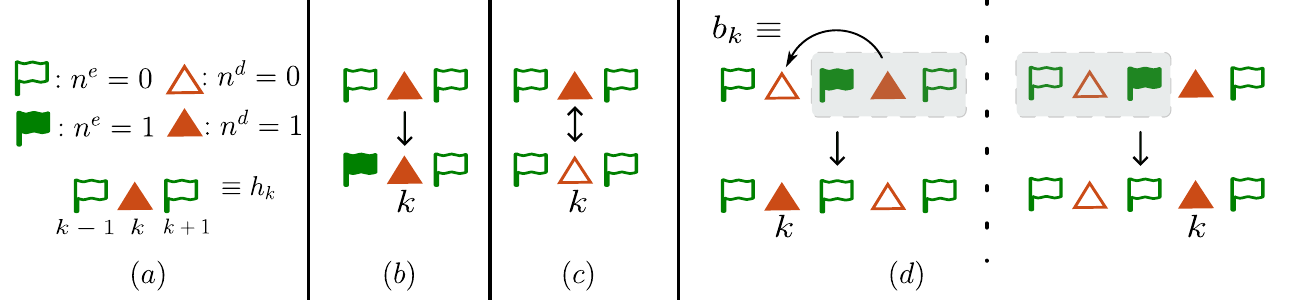}}%
}\hfill
\caption{Schematics for unheralded defect dynamics: (a) The key for the symbols is provided here. The top and bottom rows in the subsequent pictures show the initial and final configurations respectively. The dynamics is invariant under reflection symmetry about any site; hence, we do not show equivalent configurations that can be obtained as mirror images of processes shown here. (b) Erasure noise processes that remove an unheralded defect by adding flags. (c) Non-erasure noise processes flip the state of the defect without changing the flag configurations. (d) The correction protocol generates an unheralded defect when it acts on the 5-site configuration defined by $\hat{b}$. The dotted rectangle shows the triplet of sites that goes through the correction step.}\label{fig:non-erasure-scheme}
\end{figure}

The part of the noise channel that creates the defects with flags removes $h_k$ at a constant rate, resulting in 
\eq{eq:h-fe}{\frac{d h_k}{dt}\bigg|_{\text{heralded noise}} = -2\eta f_e\langle h_k\rangle.}
This corresponds to processes that remove $h_k$ by adding flags to either side of the defect at a rate $\eta f_e$ (see Fig.~\ref{fig:non-erasure-scheme}b). The unheralded part of the noise channel leads to
\eq{eq:h-non-fe}{\frac{d h_k}{dt}\bigg|_{\text{unheralded noise}} = 2\eta(1-f_e) [\ \langle (1-n^e_{k-1})(1-n^d_k)(1-n^e_{k+1})\rangle - \langle h_k\rangle].}
These processes are schematically shown in the Fig.~\ref{fig:non-erasure-scheme}c. We simplify this exact equation by assuming an uncorrelated background density of erasure flags, such that $\langle n^e_{k-1} n^e_{k+1}\rangle\sim(n^e)^2$. Using this mean-field approximation in combination with Eqs.~\eqref{eq:h-fe} and~\eqref{eq:h-non-fe}, we obtain the approximate dynamics:
\eq{}{\frac{d}{dt}h\bigg |_{\text{noise}} = 2\eta(1-f_e)(1-n^e)^2\ -\  2\eta(2-f_e) h.}

Additionally, the correction process itself generates unheralded defects by shortening those erasure strings that have a defect at only one endpoint. The exact dynamics of unheralded defects under the correction channel can be obtained as
\eq{eq:h-corr}{\eqsp{\frac{d h_k}{dt}\bigg|_{\text{correction}} &= 2\gamma \langle \ (1-n^e_{k-1})(1-n^d_k)n^e_{k+1}n^d_{k+2}(1-n^e_{k+3})\ \rangle + 2\gamma \langle (1-n^e_{k-1}) n^d_k n^e_{k+1} (1-n^d_{k+2})(1-n^e_{k+3})  \rangle \\
&:=2\gamma \langle b^R\rangle + 2\gamma \langle b^L\rangle,}}
where we have used invariance under translation symmetry to simplify the expression. The reflection symmetry of the dynamics implies that configurations hosting a defect on the left or the right side of the erasure will have the same value in the steady state. Hence, we define the relevant five-site correlation as $\hat{b}^R = \hat{b}^L=\hat{b}$. The configurations that create the unheralded defects as a result of the correction step are shown in Fig.~\ref{fig:non-erasure-scheme}d. The time-evolution depends on the correlation between fives sites, defined by the operator $\hat{b}_k$, illustrated in the first configuration in Fig.~\ref{fig:non-erasure-scheme}d.  We now simplify this exact equation by only considering the dominant processes that contribute to $b_k$, while working within the mean-field approximation discussed above.

In particular, let us focus on the parameter regime $\eta\ll\gamma$ and $f_e\sim1$. The erasure noise channel generates the configuration $b$ by adding a flag next to an isolated unheralded defect. This channel can also reduce the density of $b$ by adding flags and converting unheralded defects into heralded defects. The resulting dynamics are approximated as 
\eq{eq:b-1}{\frac{db}{dt}\bigg|_1 = \eta f_e (h(1-h) -2b).}
The non-erasure part of the noise channel can generate configurations $b$ by adding defects to an existing isolated flag. Additionally, it can decrease the density of $b$ by displacing the defect from the erasure flag. The corresponding rate equation is given by
\eq{eq:b-2}{\frac{db}{dt}\bigg|_2 = \eta (1-f_e) (n^e(1-n^e)^2 -2b).}
Finally, the correction channel converts the configuration $b$ into unheralded defects as mentioned before. This is captured by
\eq{eq:b-3}{\frac{db}{dt}\bigg|_3 = -2\gamma b.}
Note that higher order correction processes can also lead to the creation of $b$ configurations. One such process corresponds to the shortening of a longer erasure string with only one defect at its endpoint. Here, we neglect such terms since they give sub-leading contributions when $\eta\ll\gamma$. Combining the contributions from Eqs.~\eqref{eq:b-1},~\eqref{eq:b-2}, and~\eqref{eq:b-3}, we obtain the steady state density of $b$ as:
\eq{}{b=\frac{\eta f_e}{2\gamma} \ h(1-h) + \frac{\eta(1-f_e)}{2\gamma}n^e(1-n^e)^2,}
where we have dropped $\mathcal{O}\left(\eta/\gamma\right)^2$ terms. 

\begin{figure}
\subfloat[\label{subfig:hp-a}]{%
  \includegraphics[width=0.33\columnwidth]{{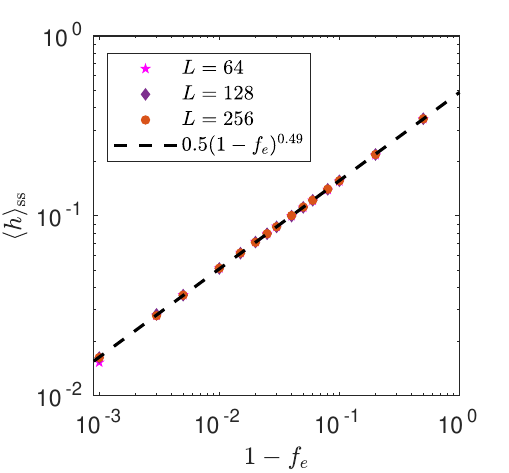}}%
}\hfill
\subfloat[\label{subfig:hp-b}]{%
  \includegraphics[width=0.33\columnwidth]{{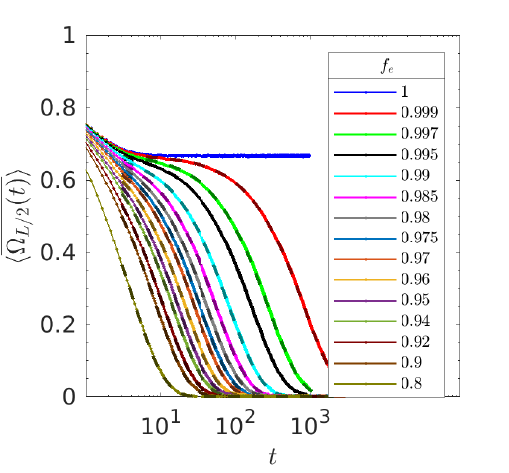}}%
}\hfill
\subfloat[\label{subfig:hp-c}]{%
  \includegraphics[width=0.33\columnwidth]{{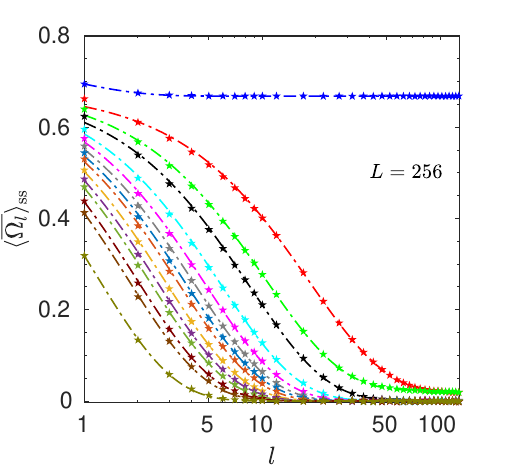}}%
}\hfill
\caption{(a) The steady state density of unheralded defects $h$ as a function of the non-erasure noise fraction $1-f_e$. The data is shown for total noise rate $\eta=0.3$ and varying system sizes. The dashed line indicates that the data for multiple system sizes matches well with $\frac{1}{2}\sqrt{1-f_e}$. (b) Estimation of $\tau_\Omega$: The string order $\Omega$ as a function of time for varying values of the erasure fraction $f_e$ and the noise rate $\eta/\gamma=0.3$. The data points obtained using Monte Carlo simulations are fitted to the function $ae^{-t/\tau_\Omega}$ (dashed curves). (c) Estimation of $\xi_\Omega$: the steady state value of the string order $\Omega_l$ is plotted as a function of the length of the string. The data is fitted to the functional form $\tilde{a}e^{-(l-1)/\xi_{\Omega}}+c$ (shown using dotted lines of same color). The variation of the time-scale $\tau_\Omega$ and length-scale $\xi_\Omega$ as a function of $f_e$ is shown in Fig.~3 in the main text.}
\end{figure}

The complete mean-field rate equation for the unheralded defects $h$ can now be obtained by adding up contributions from Eqs.~\eqref{eq:h-fe},~\eqref{eq:h-non-fe}, and~\eqref{eq:h-corr}, resulting in:
\eq{eq:h-all}{\frac{dh}{dt} = 2\eta(1-f_e)(1+n^e)(1-n^e)^2-4\eta(1-f_e)h-2\eta f_eh^2.}
Notice that the leading $\mathcal{O}(f_eh)$ term in Eq.~\eqref{eq:h-fe} is exactly cancelled by terms of the same order that result from the correction of $b$. This encodes the fact that, when the correction rate is large enough, the unheralded defects removed by the erasure noise are added back to the system at the same rate by the correction channel. Hence, we are left with a linear in $h$ term that has a smaller $\mathcal{O}(1-f_e)$ coefficient. The resulting steady state density of the unheralded defects obtained from Eq.~\eqref{eq:h-all} is
\eq{}{h=\frac{f_e-1+\sqrt{1-f_e}}{f_e}\sim\sqrt{1-f_e},}
where we have assumed that $n^e\ll1$ for parameters that are well inside the ordered phase ($\eta\ll\gamma,f_e\sim1$). The numerical values of the steady state density of unheralded defects obtained using the Monte Carlo simulations for $\eta/\gamma=0.3$ are shown in Fig.~\ref{subfig:hp-a}. The density grows $\propto \sqrt{1-f_e}$, as predicted by the mean-field equation that only considers dominant higher order processes.

\begin{figure}
\subfloat[\label{subfig:sto-fss-a}]{%
  \includegraphics[width=0.6\columnwidth]{{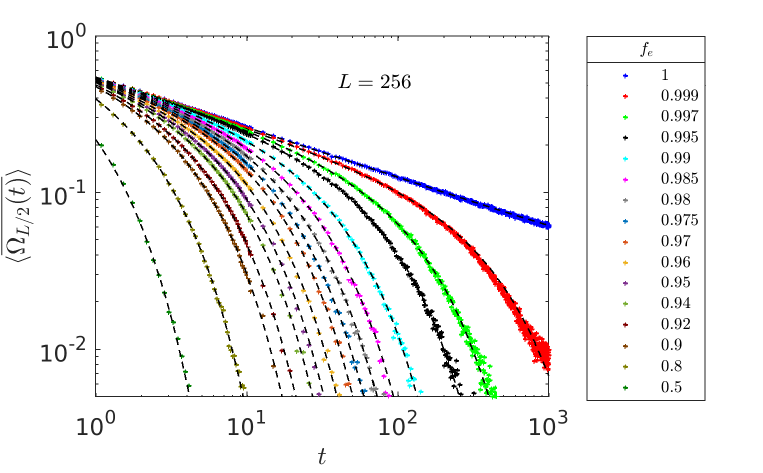}}%
}\hfill
\subfloat[\label{subfig:sto-fss-b}]{%
  \includegraphics[width=0.4\columnwidth]{{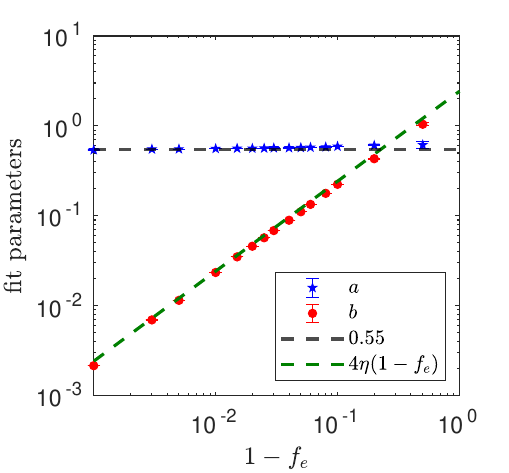}}%
}\hfill
\caption{Finite time signature of criticality: (a) The string order is plotted as a function of time for parameters close to the critical point: $\eta/\gamma=\eta_c$ and for various values of the heralding fraction $f_e$. The datapoints are obtained using Monte Carlo simulations for an $L=256$ sized system initialized in the cluster state. The dashed line show the best-fits to $at^{-\delta_\Omega}e^{-bt}$ where $a$ and $b$ are fit parameters and $\delta_\Omega=0.32$ is the order-parameter decay exponent. (b) The fit parameter $a$ takes a constant value close to the pre-factor obtained for the $f_e=1$ case (0.55). The parameter $b$ matches well with $4\eta(1-f_e)$ which corresponds to the rate of decay of $\Omega$ in the presence of unheralded $Z$ noise-channel with a rate $\eta(1-f_e)$. This captures the cross-over from an initial critical power law decay $t^{-\delta_{\Omega}}$ to a finite-rate exponential decay $e^{-4\eta(1-f_e)t}$ that is dominated by the presence of unheralded defects.}
\end{figure}

Now, we estimate the steady state value of the string order in subregion of length $l$. When the noise rate is low $\eta\ll\gamma$ and the fraction of unheralded errors is small $(f_e\sim1)$, we expect that there will be two contributions to the density of stabilizer defects. The defects that are confined due to the flags will give a finite contribution to the string order that is independent of the string length $l$. In contrast, stabilizer defects that are not heralded by flags (previously denoted by $h$) will remain deconfined. These uncorrelated defects destroy the string order of longer strings in the steady state. We estimate the string order $\Omega_l$ in terms of the unheralded defect density as 
\eq{}{\Omega_l = (1-2h)^l=e^{l\ \log{(1-2h)}} \sim e^{-2h \ l}:=e^{-l/\xi_{\Omega}},}
where we have approximated $\log{(1-2h)}\sim -2h$ for a small value of the unheralded defect density. Using this result, we estimate the string-order length-scale $\xi_\Omega\propto1/h\propto(1-f_e)^{-1/2}$. The estimated variation of $\xi_\Omega$ as a function of $f_e$ matches well with the numerical results reported in Fig.~\ref{subfig:hp-c} and Fig.~3c in the main text. Finally, in Figs.~\ref{subfig:sto-fss-a} and~\ref{subfig:sto-fss-b} we show the finite-time behavior of the string order, which displays a cross-over from an initial power law decay (dictated by $\delta_\Omega$, the order-parameter decay exponent) to an exponential decay (dictated by unheralded errors), with the cross-over occurring at a time-scale $\sim \frac{1}{\eta(1-f_e)}$.


\section{Alternative approach to stabilizing string-order}

Here, we present an alternative method for stabilizing a mixed state SPT phase with finite string order that does not rely on biased erasure noise. Instead, this mechanism uses fixed boundary conditions and endows the defects with chiral motion that effectively confines the stabilizer defects away from the boundaries of the system.

Consider a one-dimensional system consisting of $2L$ qubits and with open boundary conditions. The qubits are subjected to a strongly-symmetric noise channel, modeled by the jump operators
\eq{}{L_{\eta,j}=\sqrt{\eta}\ Z_j \qquad \text{for }\ j=3,4,\ldots,2L-2.}
We assume that the qubits located near the left (site $1$ and $2$) and the right (site $2L-1$ and $2L$) boundaries of the system are immune to this noise. This enforces the constraint that the stabilizer defects generated by the noise channel are not allowed to leak through the boundary. 

We design our correction protocol as follows: the defects undergo biased diffusion, hopping towards the center of the chain with a rate $\frac{1}{2}\gamma(1+\mu)$ and away from the center with a lower rate $\frac{1}{2}\gamma(1-\mu)$. Here, $\gamma$ is overall rate of correction and $0\leq \mu\leq 1$ parameterizes the directional bias. This effectively generates a diffusive motion of the defects that is biased towards the center of the chain. The jump operators modeling the motion in the left direction are given by
\eq{}{
\eqsp{
L_{l,j} = \begin{dcases}
    \sqrt{\frac{\gamma(1-\mu)}{2}}\ Z_j\frac{1-S_{j+1}}{2} \qquad &\text{if  }\ j= 3,4,\ldots,L \\
    \sqrt{\frac{\gamma(1+\mu)}{2}}\ Z_j\frac{1-S_{j+1}}{2} \qquad &\text{if  }\ j= L+1,L+2,\ldots,2L-2.
\end{dcases}
}
}
Similarly, the motion in the right direction is modeled using
\eq{}{
\eqsp{
L_{r,j} = \begin{dcases}
    \sqrt{\frac{\gamma(1+\mu)}{2}}\ Z_j \frac{1-S_{j-1}}{2} \qquad &\text{if  }\ j= 3,4,\ldots,L\\
    \sqrt{\frac{\gamma(1-\mu)}{2}}\ Z_j \frac{1-S_{j-1}}{2} \qquad &\text{if  }\ j= L+1,L+2,\ldots,2L-2.
\end{dcases}
}
}
Since the noise channel is inactive at the boundary, our correction channel does not act on the first or last unit cells of the system.   

The combined dynamics of the system under the noise and the correction channels is governed by the Lindblad equation
\eq{}{\frac{d\rho}{dt} = \sum_{\alpha\in\{\eta,l,r\}}\sum_{j=3}^{2L-2} L_{\alpha,j}\rho L_{\alpha,j}^\dagger - \frac{1}{2}\{L_{\alpha,j}^\dagger L_{\alpha,j} \ ,\rho\}.}
The absence of dephasing noise on the boundary qubits, along with the choice of boundary correction-operators, implies that the stabilizers $S_1=X_{2L}Z_1X_{2}$ and $S_{2L} = X_{2L-1}Z_{2L}X_1$ are conserved under this dynamics. Additionally, the jump operators in the bulk preserve the parity of the number of defects, ensuring that the operators
\eq{eq:chiral-conserve}{\prod_{i=1}^{L-1}S_{2i},\qquad\text{and}\qquad \prod_{i=2}^{L}S_{2i-1}}
are also conserved under this dynamics.  This ensures that the end-to-end string order obeys $\langle \Omega_{1, L-1} \rangle = 1 $.

\begin{figure}
\subfloat[\label{subfig:chiral-a}]{%
  \includegraphics[width=0.33\columnwidth]{{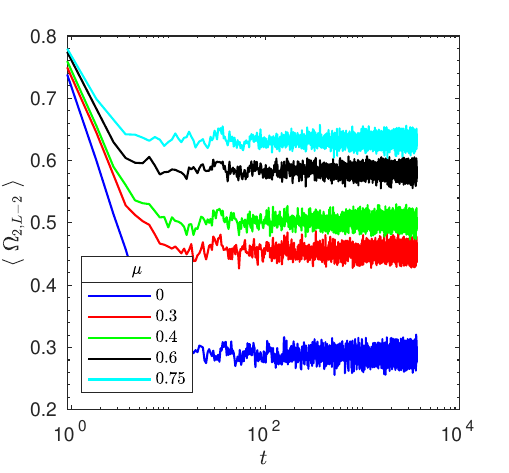}}}\hfill
\subfloat[\label{subfig:chiral-b}]{%
  \includegraphics[width=0.33\columnwidth]{{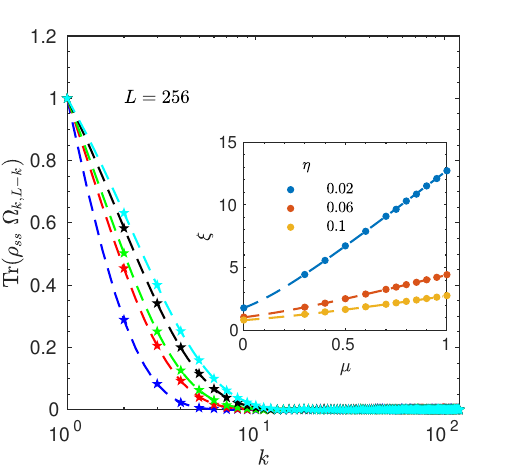}}}\hfill
\subfloat[\label{subfig:chiral-c}]{%
  \includegraphics[width=0.33\columnwidth]{{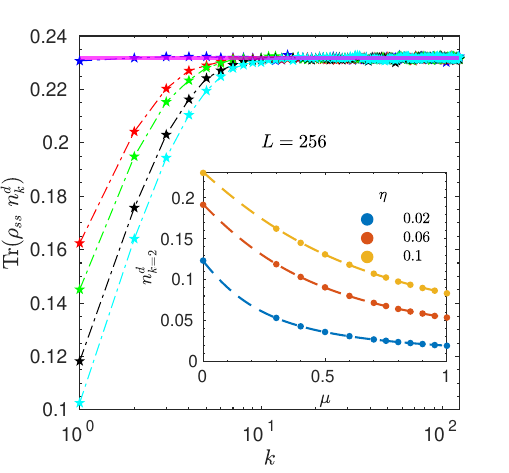}}}\hfill 
\caption{Stabilization of string order using biased motion: (a) the string order parameter $\Omega_{2,L-2}:=X_3Z_4 Z_6\ldots Z_{2L-4}X_{2L-3}$ is plotted as a function of time for varying values of the drift parameter $\mu$. The data is shown for $L=256$ sites per sub-lattice and noise rate $\eta/\gamma=0.1$. (b) The steady state value of the string order parameter as a function of the location of the left end-point $k$. The midpoint of the string is at the center of the system. The dashed lines show the best fit to an exponential function $e^{-(k-1)/\xi}$. The data-points in the inset show the length-scale $\xi$ obtained from the fits as a function of $\mu$ for different values of $\eta$. The values match well with the analytical prediction in Eq.~\eqref{eq:chiral-xi} shown by the dashed curve. (c) The average number of stabilizer defects in the steady state at each site is plotted as function of the site-location. The number of defects per site in the bulk approaches the predicted value (Eq.~\eqref{eq:chiral-nd-bulk}) shown using the magenta line. The dotted lines joining the numerical data points are a guide to the eye. The variation of the number of defects at the boundary sites is plotted in the inset as a function of $\mu$. The data-points match well with the predicted curve (Eq.~\eqref{eq:chiral-nd-2}), shown by dashed lines of same color. The numerical results are obtained by Monte Carlo simulations of the Lindblad master equation in the population basis of the stabilizers on even sites (the approximations for the heralded noise model discussed in Sec.~\ref{sec:MC} are valid for this model as well). The steady state results in the plots are evaluated after $4\times 10^3$ MC sweeps by averaging over $10^4$ independent Monte Carlo samples initialized in the cluster state.}
\end{figure}

The chiral dynamics pushes the stabilizer defects towards the center of the system, bringing them close together and allowing for local correction. Let us first focus on sites in the bulk of the system that are away from the center and from the endpoints (consider left half of the chain, calculations work similarly for right half due to reflection symmetry). The exact time evolution of the density of defects at such sites $k$ is governed by 
\eq{}{\frac{d}{dt}\langle n^d_k\rangle = 2\eta(1-2\langle n^d_k\rangle) -\gamma \langle n^d_k\rangle +\frac{\gamma}{2}\left[ (1+\mu)\langle n^d_{k-2}(1-2n^d_k) \rangle+ (1-\mu) \langle n^d_{k+2}(1-2n^d_k)\rangle \right]  ,}
where $\langle\ .\ \rangle$ denotes the expectation value of the observable in the time evolved mixed-state. A mean-field approximation $\langle n^d_kn^d_{k+2}\rangle\sim (n^d)^2$ simplifies this exact expression to a rate equation, given by
\eq{}{\frac{d n^d}{dt} = -2(\gamma+\eta)(n^d)^{2} + 2\eta (1-n^d)^2.}
This expression captures the effective dynamics wherein defects are created in pairs at originally unoccupied adjacent sites at a rate $2\eta$ and are removed at a rate $2(\gamma+\eta)$. The uniform density implies that the current in and out of the region around the site are the same and hence the bias parameter $\mu$ does not affect the overall density. In the steady state, the density of defects in the bulk of the system saturates to 
\eq{eq:chiral-nd-bulk}{n^d_{ss}=-\frac{\eta}{\gamma}+ \sqrt{\frac{\eta}{\gamma}\left(1+ \frac{\eta}{\gamma}\right)}.}
The sites at the boundary, on the other hand, only have defects entering from one side. For example, the number of defects at site-2 evolve as
\eq{}{\frac{d \langle n^d_2\rangle}{dt} = \eta \langle1-2n^d_2\rangle -\gamma\frac{1+\mu}{2} n^d_2 + \gamma\frac{1-\mu}{2}\langle(1-2n^d_2)n^d_4\rangle.}
Using a mean field approximation where $\langle n^d_4 n^d_2\rangle\sim (n^d_2)^2$ , the steady state density at the left endpoint is given by
\eq{eq:chiral-nd-2}{n^d_2 = \frac{2\eta+\gamma\mu - \sqrt{4\gamma\eta+4\eta^2+\gamma^2\mu^2}}{2\gamma(\mu-1)}.}
The numerical values for the defect density obtained using Monte Carlo simulations matches well with these analytical estimates (as shown in Fig.~\ref{subfig:chiral-c}).

We analyze the SPT order using the string order parameter $\Omega_{k,L-k}=\prod_{i=k}^{L-k} S_{2i}$ (see Eq.~\eqref{eq:string_order}). This string is centered at the midpoint of the system, with $k$ denoting the distance between the string endpoint and the system boundary measured in terms of unit cells. The exact time evolution is given by
\eq{eq:omega-chiral}{\eqsp{\frac{d\langle \Omega_{k,L-k}\rangle}{dt} =& -(4\eta+2\gamma)\langle\Omega_{k,L-k}\rangle
+ \frac{\gamma(1+\mu)}{2}\langle \Omega_{k-1,L-k} + \Omega_{k,L-k+1}\rangle \\
&+ \frac{\gamma(1-\mu)}{2} \langle \Omega_{k+1,L-k} +  \Omega_{k,L-k-1}\rangle. 
}}
We assume that the string order varies smoothly as a function of the position of the endpoints of the string ($x:=-a+ka$) which leads to 
\eq{}{\eqsp{\langle \Omega_{k-1,L-k}\rangle = \langle\Omega_{k,L-k+1}\rangle &\ \sim\ \Omega(x) - \frac{a}{2}\frac{d\Omega}{dx} + \frac{(a/2)^2}{2}\frac{d^2\Omega}{dx^2} \\ 
\langle \Omega_{k+1,L-k}\rangle =  \langle\Omega_{k,L-k-1}\rangle &\ \sim \ \Omega(x) + \frac{a}{2}\frac{d\Omega}{dx} + \frac{(a/2)^2}{2}\frac{d^2\Omega}{dx^2}  },}
where $a$ is the size of unit cell, and where the additional factor of one-half indicates that the length of the string is only modified from one side. The first set of equalities in the equation above result from the invariance of the dynamics with respect to the reflection symmetry about the center of the chain. 
Using this continuum approximation, along with Eq.~\eqref{eq:omega-chiral}, we obtain the steady state equation for the string order as
\eq{}{\frac{(a/2)^2\gamma}{2}\frac{d^2\Omega}{dx^2} -\gamma\mu\frac{a}{2}\frac{d\Omega}{dx} -2\eta\ \Omega =0.}
Notice that $\Omega(x=0)$ can be written as $\prod_{i=1}^{L-1}S_{2i}$. As a result of the dynamics that does not allow stabilizer defects to leak through the boundary (Eq.~\eqref{eq:chiral-conserve}), this operator is conserved and takes the value $1$ if the system is initialized in the cluster state. Hence, by definition $\Omega$ across the entire string is equal to $1$ for all times. Using this boundary condition, we get the steady state string order (setting $a=1$) to be
\eq{eq:chiral-xi}{\Omega_{ss}(x) = \text{exp}\left[- 2 x \left( \sqrt{\frac{4\eta}{\gamma}+\mu^2}-\mu\right)  \right]:=e^{-x/\xi}.}
This implies that there is a finite steady-state string order when both the left and right endpoints of the string are within a distance $\xi$ of the respective system-boundaries. Since the length of such a string is equal to $L-2x$, we have a steady state that has a boundary sensitive long-range string order. In Fig.~(\ref{subfig:chiral-a}), we numerically verify the long-time string order for a system initialized in the cluster state. Additionally, the analytical mean-field estimate for the length scale $\xi$ also matches well with the numerical data for a wide range of parameters (see Fig.~(\ref{subfig:chiral-b})).

In summary, the fixed boundary condition (i.e., requiring the absence of $Z$ noise at the endpoints) ensures that the parity of defects in the whole system is conserved. The additional chiral motion induced by the correction protocol creates a region of size $\xi$ near the boundary of the system that has a lower number of  defects compared to the bulk of the system. This results in a finite value of the string order parameter when the endpoints of the string are close to the boundary, resulting in nontrivial steady-state SPT order.




\end{document}